\documentclass[12pt]{article}%
\usepackage{amsmath}
\usepackage{float}
\usepackage{theorem}
\usepackage{graphicx}
\usepackage{amsfonts}
\usepackage{amssymb}%
\providecommand{\U}[1]{\protect\rule{.1in}{.1in}}
\providecommand{\U}[1]{\protect\rule{.1in}{.1in}}
\providecommand{\U}[1]{\protect\rule{.1in}{.1in}}
\providecommand{\U}[1]{\protect\rule{.1in}{.1in}}
\providecommand{\U}[1]{\protect\rule{.1in}{.1in}}
\providecommand{\U}[1]{\protect\rule{.1in}{.1in}}
\providecommand{\U}[1]{\protect\rule{.1in}{.1in}}
\providecommand{\U}[1]{\protect\rule{.1in}{.1in}}
\providecommand{\U}[1]{\protect\rule{.1in}{.1in}}
\providecommand{\U}[1]{\protect\rule{.1in}{.1in}}
\providecommand{\U}[1]{\protect\rule{.1in}{.1in}}
\providecommand{\U}[1]{\protect\rule{.1in}{.1in}}
\providecommand{\U}[1]{\protect\rule{.1in}{.1in}}
\providecommand{\U}[1]{\protect\rule{.1in}{.1in}}
\providecommand{\U}[1]{\protect\rule{.1in}{.1in}}
\providecommand{\U}[1]{\protect\rule{.1in}{.1in}}
\providecommand{\U}[1]{\protect\rule{.1in}{.1in}}
\providecommand{\U}[1]{\protect\rule{.1in}{.1in}}
\providecommand{\U}[1]{\protect\rule{.1in}{.1in}}
\providecommand{\U}[1]{\protect\rule{.1in}{.1in}}
\providecommand{\U}[1]{\protect\rule{.1in}{.1in}}
\theorembodyfont{\upshape}
\newtheorem{theorem}{Theorem}

\newtheorem{claim}[theorem]{Claim}

\newtheorem{assumption}[theorem]{Assumption}

\newtheorem{corollary}[theorem]{Corollary}

\newtheorem{definition}[theorem]{Definition}
\newtheorem{example}[theorem]{Example}

\newtheorem{lemma}[theorem]{Lemma}
\newtheorem{notation}[theorem]{Notation}

\newtheorem{remark}[theorem]{Remark}

\newenvironment{proof}[1][Proof]{\textbf{#1.} }{\ \rule{0.5em}{0.5em}}
\advance\topmargin by -0.75truein
\advance\textheight by 1.25truein
\begin{document}

\begin{center}
{\LARGE On modeling the TSP as a polynomial-sized linear program:
Escaping/Side-stepping the \textquotedblleft barriers\textquotedblright%
}\bigskip

Moustapha Diaby

OPIM Department; University of Connecticut; Storrs, CT 06268\\[0pt]%
moustapha.diaby@uconn.edu\medskip{\Large \medskip}

Mark H. Karwan

Department of Industrial and Systems Engineering; SUNY at Buffalo; Amherst, NY
14260\\[0pt]mkarwan@buffalo.edu\medskip{\Large \medskip}

Lei Sun

Department of Industrial and Systems Engineering; SUNY at Buffalo; Amherst, NY
14260\\[0pt]leisun@buffalo.edu\medskip\medskip
\end{center}

\textsl{Abstract:}{\small \ In view of the \textit{extended formulations}
(EFs) developments, we bring a clarification in this paper to the question of
whether it is possible to model an NP-Complete problem as a polynomial-sized
linear program. For the sake of simplicity of exposition, we focus on the
traveling salesman problem (TSP). We show that a finding that there exists no
polynomial-sized EF of\ \textquotedblleft\textit{the} TSP
polytope\textquotedblright\ does not (necessarily) imply that it is
\textquotedblleft impossible\textquotedblright\ for a polynomial-sized linear
program to solve the TSP optimization problem. We show that under appropriate
conditions the TSP optimization problem can be solved without recourse to the
traditional city-to-city (\textquotedblleft travel leg\textquotedblright)}
{\small variables,\ thereby side-stepping/\textquotedblleft escaping
from\textquotedblright\ \textquotedblleft\textit{the} TSP
polytope\textquotedblright\ and hence, the \textquotedblleft
barriers.\textquotedblright\ Some illustrative examples are discussed.
}\bigskip\medskip

\textsl{Keywords:}\textbf{\ }{\small Combinatorial Optimization; Computational
Complexity; Linear Programming; Linear Assignment Problem; Traveling Salesman
Problem.\bigskip}

\section{Introduction\label{Introduction_Section}}

A fundamental issue for modern mathematics in general has been the question of
decidability/computability (Hilbert [1901]; Turing [1936]). The
decidability/computability question was reduced (in essence) to that of the
equality/non-equality of the computational complexity classes
\textquotedblleft P\textquotedblright\ and \textquotedblleft
NP\textquotedblright\ (\textquotedblleft P vs. NP\textquotedblright\ problem)
in the early 1970's (Cook [1971]). Roughly, the \textquotedblleft
NP\textquotedblright\ class of problems are such that checking the correctness
of a given solution to them (respectively) is \textquotedblleft
easy.\textquotedblright\ The \textquotedblleft P\textquotedblright\ class
consists of \textquotedblleft NP\textquotedblright\ problems which are known
to also be \textquotedblleft easy\textquotedblright\ to solve (i.e., for which
polynomial-time algorithms are known). Hence, \textquotedblleft
P\textquotedblright\ is a subclass of \textquotedblleft NP.\textquotedblright%
\ Within the \textquotedblleft NP\textquotedblright\ class, the so-called
NP-Complete (NPC; see Garey and Johnson [1979], among others) problems play a
central role in the sense that: (1) they are polynomially reducible to each
other among themselves; and (2) every other problem in \textquotedblleft
NP\textquotedblright\ is polynomially reducible to any of them. The essence of
the \textquotedblleft P vs. NP\textquotedblright\ problem is the question of
whether or not \textquotedblleft P\textquotedblright\ is a proper subclass of
\textquotedblleft NP,\textquotedblright\ or in other words, whether or not
there exists a polynomial time procedure for solving any of the NPC problems.

In the field of mathematical programming, NPC problems are generally analyzed
under the umbrella of combinatorial optimization. Hence, with the advent of
polynomial-time algorithms for solving linear programs (LPs) in the early
1980's (Khachiyan [1979]; Karmarkar [1984]), the \textquotedblleft P vs.
NP\textquotedblright\ problem was reduced in part (for operations research
(OR) in particular) to the question of whether or not any of the NP-Complete
combinatorial optimization problems (COPs) can be modeled as a
polynomial-sized LP. The challenge in this sense is to develop a set of linear
constraints in such a way that the extreme points of the resulting LP are in a
one-to-one correspondence with the combinatorial configuration (e.g.
\textquotedblleft tours,\textquotedblright\ \textquotedblleft
partitions,\textquotedblright\ \textquotedblleft covers,\textquotedblright%
\ etc.) for the COP at hand. The NPC problem which has been used in research
focused on this issue in general has been the traveling salesman problem
(TSP). NP-Completeness pertains to decision problems. However, throughout the
remainder of this paper, we will use the TSP optimization problem (which is
NP-Hard; see Garey and Johnson [1979], among others) instead of the TSP
decision problem. Our reasons for doing this are convenience (since the
framework of our exposition is a mathematical programming one) and the fact
that the TSP optimization problem subsumes the TSP decision problem.

The first polynomial-sized LP models proposed for solving the TSP date back to
the mid-1980's (Swart [1986/1987]). We do not know of any details of these
models. However, the consensus of the research communities is that their
validities were refuted by a negative result (\textquotedblleft
barrier\textquotedblright) developed in the early 1990's (Yannakakis [1991]).
The most prominent conclusion of the Yannakakis [1991] work is that
\textquotedblleft\textit{the} TSP polytope\textquotedblright\ does not have a
polynomial-sized extended formulation (EF) which is symmetric. By
\textquotedblleft\textit{the} TSP polytope,\textquotedblright\ we mean the
polytope described using the traditional/standard/\textquotedblleft
natural\textquotedblright/\textquotedblleft travel leg\textquotedblright%
/\textquotedblleft city-to-city\textquotedblright\ variables for the TSP.
Since these Yannakakis [1991] developments, making inferences about possible
LP models for COPs by analyzing EFs of their \textit{natural} polytopes has
become a standard practice. Recently, Fiorini et al. [2015] strengthened the
Yannakakis \textit{barrier} into one which no longer has the symmetry caveat.
That work has been, in turn, seminal for many of the other recent developments
(e.g. Faenza et al. [2021], Aprile [2022], and Kwan et al. [2022]).

The purpose of this paper is not to dispute the validity of the
\textit{barriers} themselves. Rather, our objective is to show that the
non-existence of a polynomial-sized integral extended formulation of
\textquotedblleft\textit{the} TSP polytope\textquotedblright\ does not imply
that it is \textquotedblleft impossible\textquotedblright\ for a
polynomial-sized linear program to solve the TSP optimization problem. The
importance of this comes from the significance of NPC problems for both
practice and theory in many fields of study and the fact that,  in the
literature, there have been assertions based on developments for
\textquotedblleft\textit{the} TSP polytope\textquotedblright\ that it is
\textquotedblleft impossible\textquotedblright\ to formulate any of these
problems as a linear program.

An optimization problem (using mathematical programming) is defined by both
its constraints and its objective function. An oddity of the TSP (and
incidentally, of many of the other NP-Complete problems as well) is its
relation to the linear assignment problem (LAP)/bi-partite matching problem.
Specifically, each extreme point of the LAP defined by its linear constraint
set can be interpreted as a TSP tour with the assignment of row $i$ to column
$j$ (like job $i$ assigned to person $j$) interpreted as city $i$ being the
$j^{th}$ city visited. The difficulty in this abstraction however, is that the
cost function of traveling from city to city cannot be captured using the
two-dimensional, \textit{natural} variables traditionally used to state the
constraints of the LAP. Hence, the key to being able to exploit this feature
is to extend the LAP into a higher dimension in a way that an objective
function relating to sets of joint assignments can be expressed. Also, the
numbers of variables and constraints must respectfully be polynomial, and the
model must have all-integer extreme points and project to the LAP polytope. As
far as we know, this possibility is not ruled out by any existing result as it
may be able to be accomplished without regard to the traditional city-to-city
variables of the TSP but only by using an appropriate cost function on the EF
of the LAP.

The plan of the remainder of the paper is as follows. In section
\ref{Preliminary_Section}, we give an overview of our basic notions,
definitions, and notations; discuss alternate representations of TSP tours;
and finally, briefly overview the existing \textit{barriers} and discuss some
simple misconceptions and misinterpretations in relation to them. In section
\ref{SideStepping_Section}, we provide a motivating illustrative example and
develop conditions under which \textquotedblleft\textit{the} TSP
polytope\textquotedblright\ (and hence, the \textit{barriers}) can be
side-stepped/\textquotedblleft escaped from\textquotedblright\ with respect to
the goal of developing polynomial-sized LP models for solving the TSP
optimization problem. Finally, we offer some conclusions in section
\ref{Conclusions_Section}.

\section{Preliminaries \label{Preliminary_Section}}

\subsection{Basic notions, definitions, and notations}

\begin{figure}[ptb]
\begin{center}
\includegraphics[width=.4\textwidth]{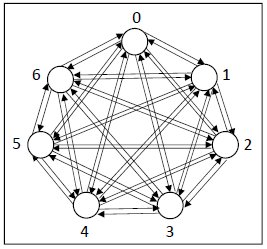}
\end{center}
\caption{Illustration of the \textit{TSP Graph}}%
\label{TSP_Graph_Figure}%
\end{figure}

Our discussions will involve two alternate abstractions of the TSP
optimization problem. We give overviews of these and their accompanying
notations in this section. The graph which underlies the first abstraction
(which we label the \textquotedblleft traditional abstraction") is illustrated
in Figure \ref{TSP_Graph_Figure}. Following the terminology commonly used in
the literature, we refer to this graph as the \textquotedblleft TSP
graph.\textquotedblright\ Each node of the \textit{TSP graph} represents a
city, whereas each arc represents a \textit{travel leg}.

\begin{figure}[ptb]
\begin{center}
\includegraphics[width=.42\textwidth]{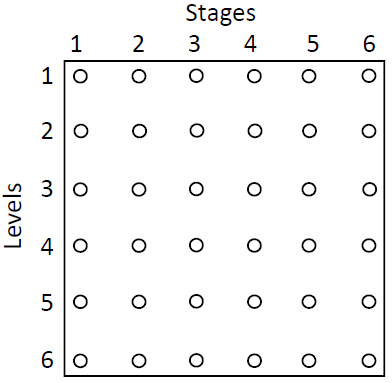}
\end{center}
\caption{Illustration of the TSP Assignment Graph (TSPAG)}%
\label{TSPAG_Figure}%
\end{figure}

The\ second abstraction we consider in this paper (which we label as the
\textquotedblleft alternate representation\textquotedblright)\ is a variant of
the \textit{time-dependent }model first proposed by Picard and Queyranne
[1978]. In this representation, city $0$\ always occupies the first and last
position in the sequence formed from a given TSP tour. In other words, we
assume that every travel begins and ends at city $0$.\ Hence in the
sequencing/permutation-finding optimization problem, the \textquotedblleft
positioning\textquotedblright\ of city $0$\ ceases to be a decision variable.
In other words, the search for an optimal TSP tour reduces to a search for a
permutation over the set of cities, minus city $0.$\ Hence, the graph which
underlies this abstraction is correctly illustrated by Figure
\ref{TSPAG_Figure}. Observe that this graph is a generic Assignment Problem
graph (see Burkard \textit{et al.} [2009]). In the context of job assignments
for example, each row of nodes of this graph (which we will refer to as a
\textquotedblleft level\textit{\textquotedblright} of the graph) could
represent a worker, and each column of nodes (which we refer to as a
\textquotedblleft stage\textit{\textquotedblright} of the graph) could
represent a task. Hence, we will refer to this graph as the \textquotedblleft
TSP Assignment Graph (TSPAG).\textquotedblright\ 

The formalization of these graphs above for a TSP with $n$ cities and the
general notation which will be used throughout the remainder of the paper are
shown below. (For convience, table-lists of all the notations used in the
paper are also provided in the Appendix to this paper.)

\begin{notation}
[General Notation]\label{General_Notation} \ \ 
\end{notation}

\begin{enumerate}
\item $n:$ \ Number of TSP cities.

\item $m:=n-1$ \ \ (Numbers of \textit{levels} and \textit{stages}
respectively, of the TSPAG).

\item $C:=\left\{  0,\ldots,m\right\}  :$ Set of cities.

\item $\forall(i,j)\in C^{2},$ $d_{ij}:$ \ Cost of traveling from city $i$ to
city $j$.

\item $\ell:=n\times m$ \ \ (Number of arcs in the \textit{TSP graph}).

\item $\mathcal{P}_{n}:$ \textquotedblleft\textit{the} TSP
polytope\textquotedblright\ defined over $n$ cities.

\item $L:=\{1,\ldots,m\}$ \ \ (Index set for the \textit{levels} of the TSPAG).

\item $S:=\{1,\ldots,m\}$ \ \ (Index set for the \textit{stages} of the TSPAG).

\item $N:=\left\{  \lceil l,s\rfloor:\text{\textquotedblleft}%
(l,s)\text{\textquotedblright},\text{ }(\forall l\in L),\text{ }(\forall s\in
S)\right\}  $ \ (There is no \textquotedblleft ceiling\textquotedblright\ or
\textquotedblleft floor\textquotedblright\ meaning attached to
\textquotedblleft$\lceil$\textquotedblright\ and \textquotedblleft$\rfloor
$\textquotedblright.).

\item $\mathcal{A}_{n}:$ The LAP polytope defined over the TSPAG associated
with a $n$-city TSP.

\item \label{GN1}$(A)^{t}:$ \ Transpose of matrix $A.$

\item \label{GN2}$Ext(B):$ \ Set of extreme points of polyhedron $B$.

\item \label{GN2.5}$\pi_{x}(B):$ \ Projection of polyhedron $B$ on the space
of vector $x$.

\item \label{GN3}$\mathbb{R}:$ \ Set of real numbers.

\item \label{GN3.2}$\mathbb{R}_{\mathbb{\geq}0}:$ \ Set of nonnegative real numbers.

\item \label{GN3.5}$\mathbb{N}:$ \ Set of natural numbers (excluding
\textquotedblleft$0$\textquotedblright).
\end{enumerate}

\subsection{Alternate representations of TSP tours}

\subsubsection{\textquotedblleft\textit{The} TSP polytope\textquotedblright%
\label{TSP_Polytope_SubSection}}

\begin{figure}[ptb]
\begin{center}
\includegraphics[width=.4\textwidth]{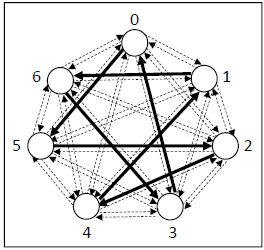}
\end{center}
\caption{Illustration of a \textit{Hamiltonian Cycle}}%
\label{Hamiltonian_Cycle_Illustration}%
\end{figure}

The existing negative results (\textit{barriers}) are based on the
conventional, standard abstraction which focuses on the arcs of the
\textit{TSP graph} (i.e., the \textit{travel legs} of the TSP). The goal in
this abstraction is to select a set of arcs of the \textit{TSP graph }which
form a Hamiltonian cycle of the nodes of the graph. This is illustrated in
Figure \ref{Hamiltonian_Cycle_Illustration}. Hence, in this abstraction, a TSP
tour can be represented in terms of the TSP cities as a $(n+1)$-tuple
$h=(h_{0}=$ $0,h_{1},...,h_{n-1},h_{n}=0)$ such that $\{h_{1},...,h_{n-1}%
\}=C\backslash\{0\}$. The TSP tour represented by a given tuple $h$ can also
be represented in terms of \textit{travel legs} as $\tau(h):=\{(h_{k}%
,h_{k+1})\in h^{2},$ $k=0,...,n-1\}.$ \ 

For the purpose of a linear programming (LP) modeling, attach a vector of
variables $y\in\{0,1\}^{n(n-1)}$ to the \textit{travel legs} with $y_{ij}%
\in((i,j)\in C^{2})$ equal to $1$ if and only if there is travel from city $i$
to city $j$. The TSP tour represented by a given tuple $h$ is then modeled by
the $y$-instance, $\overline{y}^{h},$ which is such that, for each $(i,j)\in
C^{2}$, $\overline{y}_{ij}^{h}$ is equal to $1$ if and only if $(i,j)\in
\tau(h).$ \textquotedblleft\textit{The} TSP polytope\textquotedblright\ (which
we denote by $\mathcal{P}_{n}$) is defined as the polytope in the space of $y$
which has the $\overline{y}^{h}$'s as its extreme points.

Note that city $0$\ can be duplicated into a \textit{dummy} city $n,$\ so that
the condition \textquotedblleft$h_{n}=$ $0$\textquotedblright\ can be
alternatively written as \textquotedblleft$h_{n}=n.$\textquotedblright\ Hence,
$h$ represents a contrained permutation of the TSP cities, since the
\textquotedblleft$h_{0}=$ $0$\textquotedblright\ and \textquotedblleft$h_{n}=$
$0$\textquotedblright\ requirements must be enforced explicitly (including
through some classic types of inequalities (see Lawler et al. [1988])). Hence,
in other words, in an exact EF of $\mathcal{P}_{n}$,\ each extreme point of
the model must represent a constrained/restricted permutation of the TSP
cities. Hence, such EF would essentially include \textquotedblleft
side-constraints\textquotedblright\ in addition to the descriptive constraints
of the LAP polytope defined over (an Assignment-graph version of) the
\textit{TSP graph}, while -at the same time- being required to be integral.
Solutions of this \textit{TSP graph}-based LAP which comprise subtours of the
cities must also be excluded from consideration through additional
\textquotedblleft side-constraints\textquotedblright\ (such as in the
classical Dantzig et. al. [1954] model, or such as relating to the \textit{cut
polytope} discussed in Fiorini et al. [2015; section 3.2] for example). We
believe that the notorious difficulty involved in coming up with a
\textquotedblleft small\textquotedblright\ LP model of the TSP optimization
problem in which the \textit{travel leg} variables $y$ are
irredundant/non-superfluous (from an optimization perspective) and the
constraints set is an EF of $\mathcal{P}_{n}$ resides principally in these
facts. \ 

\subsubsection{LAP solution representation of TSP
tours\label{LAP_Representation_Subsection}}

A simplification of the representation described in section
\ref{TSP_Polytope_SubSection} above can be obtained by observing that under
the assumption that all travels begin and end at city $0,$\ an $m$-tuple
$p=(p_{1},...,p_{m})$ which is such that $\{p_{1},...,p_{m}\}=C\backslash
\{0\}$ is sufficient for abstracting a TSP tour. The problem of choosing from
among these $m$-tuples is most naturally modeled by the LAP defined over the
TSPAG. To specify this model, attach to the nodes of the TSPAG variables
$w\in\{0,1\}^{m^{2}}$ with $w_{ir}$ $(i\in L,$ $r\in S)$ equal to $1$ if and
only if $i$ is the $r^{th}$ city visited after city $0$. The instance of $w$,
$\overline{w}^{p},$ associated with tuple $p$ has set of positive components
$\{\overline{w}_{p_{1},1},\ldots,\overline{w}_{p_{m},m}\},$ and the sequence
of the visits in the TSP tour represented by $p$ is $0\longrightarrow
p_{1}\longrightarrow...\longrightarrow p_{m}\longrightarrow0.$ It is
well-known (see Burkhard et al. [2009]) that the polytope which has the
$\overline{w}^{p}$'s as its extreme points is the LAP/bipartite-matching
polytope in the space of $w$, which we denote by $\mathcal{A}_{n}.$ Hence,
there exists a one-to-one correspondence between the TSP tours and the extreme
points of $\mathcal{A}_{n}.$

In order for the modeling based on this alternate abstraction of TSP tours to
be complete with respect to the task of solving the TSP optimization problem,
a judicious extension of $\mathcal{A}_{n}$ in which TSP tour costs are
correctly captured through auxiliary variables must be developed. In section
$\ref{SideStepping_Section}$ of this paper, we show that such an EF would
allow for the solution of the TSP optimization problem irrespective of whether
it is polynomial-sized or not and of whether it projects to \textquotedblleft%
\textit{the} TSP polytope\textquotedblright\ ($\mathcal{P}_{n}$) or not.

We will close this section with notions which formalize the representation of
TSP tours in terms of the nodes of the TSPAG. \ 

\begin{definition}
[\textquotedblleft TSP paths\textquotedblright]\label{Path_Definitions} \ \ \ 

\begin{enumerate}
\item \label{TSP_Path_Dfn.1}We refer to a set of nodes at consecutive
\textit{stages} of the TSPAG with exactly one node at each \textit{stage} in
the set involved as a \textit{path} of the TSPAG. In other words, for
$(r,s)\in S^{2}:s>r,$ we refer to $\left\{  \lceil u_{p},p\rfloor\in N,\text{
}p=r,\ldots,s\right\}  $ as a \textit{path} of the TSPAG.

\item \label{TSP_Path_Dfn.2}We refer to a \textit{path} of the TSPAG which
(simultaneously) spans the \textit{stages} and the \textit{levels} of the
TSPAG as a \textit{TSP path} (of the TSPAG). In other words, letting $m$ be
the number of \textit{stages} of the TSPAG, we refer to $\left\{  \lceil
u_{p},p\rfloor\in N,\text{ }p=1,\ldots,m;\text{ }(p\neq q\text{ }\iff\text{
}\right.  $\newline$\left.  \text{ }u_{p}\neq u_{q})\right\}  $ as a
\textit{TSP path} (of the TSPAG). \ 
\end{enumerate}
\end{definition}

\begin{figure}[ptb]
\begin{center}
\includegraphics[width=.6\textwidth]{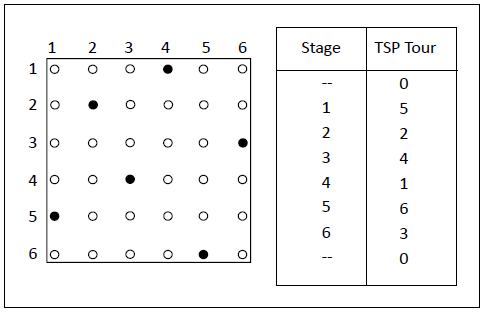}
\end{center}
\caption{Illustration of \textit{TSP paths}}%
\label{TSP_Path_Illustration}%
\end{figure}

A \textit{TSP path} is illustrated in Figure \ref{TSP_Path_Illustration}. \ 

\pagebreak

\begin{remark}
\label{Path&Tour_Correspondence_Remark} \ 

\begin{enumerate}
\item The \textit{TSP paths} of the TSPAG are in a one-to-one correspondence
with the TSP tours of the TSP.

\item Every extreme point of the Linear Assignment Problem (LAP) polytope
defined over the TSPAG ($\mathcal{A}_{n}$) corresponds to exactly one TSP tour
and vice versa. \ \ 
\end{enumerate}
\end{remark}

\subsection{Overview of the \textit{barriers }}

\subsubsection{Basic results}

\begin{definition}
[{{\textit{Extended Formulation} (EF); Yannakakis [1991]} }]\label{EF_Dfn_Std}
A polytope
\[
V:=\left\{  \left(
\begin{array}
[c]{c}%
u\\
v
\end{array}
\right)  \in\mathbb{R}^{\mu+\nu}:E^{=}u+F^{=}v=g^{=},\text{ }E^{\leq}%
u+F^{\leq}v\leq g^{\leq}\right\}
\]
is an EF of
\[
U:=\left\{  u\in\mathbb{R}^{\mu}:Au\leq b\right\}
\]
if
\[
\pi_{u}(V)=\left\{  u\in\mathbb{R}^{\mu}:\left(  \exists v\in\mathbb{R}^{\nu
}:\left(
\begin{array}
[c]{c}%
u\\
v
\end{array}
\right)  \in V\right)  \right\}  =U.
\]%
\begin{align*}
\text{(where }  &  E^{=}\in\mathbb{R}^{m^{=}\times\mu},\text{ }F^{=}%
\in\mathbb{R}^{m^{=}\times\nu},\text{ }g^{=}\in\mathbb{R}^{m^{=}},\\[0.06in]
&  E^{\leq}\in\mathbb{R}^{m^{\leq}\times\mu},\text{ }F^{=}\in\mathbb{R}%
^{m^{\leq}\times\nu},\text{ and }g^{\leq}\in\mathbb{R}^{m^{\leq}}.\text{)}%
\end{align*}

\end{definition}

\begin{definition}
[{\textquotedblleft Symmetric Polytope"; {Yannakakis [1991]}}]%
\label{Symmetry_Dfn}:

\begin{enumerate}
\item A polytope $U$ in the space of \textit{travel leg} variables
$u=(u_{ij})$ and auxiliary vaiables $v$ is symmetric if every given
permutation of the TSP nodes can be also extended to the auxiliary variables
$v$ so that $U$ remains invariant.

\item A set of linear constraints is called symmetric if its feasible space is.
\end{enumerate}
\end{definition}

\begin{theorem}
[{{\textit{Barrier}} \#1; {Yannakakis [1991]}}]\label{Barrier_1}%
\textquotedblleft The TSP polytope does not have a \textit{symmetric} EF of
subexponential size$.$\textquotedblright\ 
\end{theorem}

\begin{theorem}
[{\textit{Barrier }\#2; {Fiorini et al. [2015]}}]\label{Barrier_2}%
\textquotedblleft The extension complexity of the TSP polytope TSP(n) is
$2^{\Omega(\sqrt{n})}$.\textquotedblright\ \medskip

(This statement of \textit{Barrier \#2} can be paraphrased to make it
\textquotedblleft line up\textquotedblright\ more directly with Yannakakis'
as: \textquotedblleft The TSP polytope does not have an EF of subexponential
size.")\medskip
\end{theorem}

\subsubsection{Inferences and Misconceptions
\label{Inferences&MisconceptionsSubSubSection}}

Although the discussions in this section do not have any bearing (or only a
peripheral one at best) on the developments in this paper, we believe they
highlight the importance of the point of the paper, namely, that the finding
of the \textit{barriers} above (Theorems \ref{Barrier_1}-\ref{Barrier_2}) does
not imply that it is \textquotedblleft impossible\textquotedblright\ for a
polynomial-sized LP to solve the TSP optimization problem, as suggested in
Fiorini et al. [2015] in particular.

As indicated earlier, the purpose of this paper is not dispute the validity of
the \textit{barriers} themselves, but to show that some of the inferences that
can be/have been made from them about what is/is not possible in terms of the
LP modeling of the TSP optimization problem in general are overscoped. We will
use the equivalence between the notions of \textit{extension} and
\textit{extended formulation} discussed in Fiorini et al. [2015] for example
to provide a rough illustration of this, as follows.

\begin{claim}
[Reasoning]:\label{Reasoning_Claim} \ Let $u$ be arbitrary variables which are
disjoint from the \textit{travel leg} variables, $y$, of \textquotedblleft%
\textit{the} TSP polytope\textquotedblright\ $\mathcal{P}_{n}$. Let there be a
claim that a linear program (LP) over
\[
U:=\left\{  u\in\mathbb{R}^{\mu}:Au\leq b\right\}  \text{ \ (where }\mu
\in\mathbb{N})
\]
models the ($n$-city) TSP exactly. One would arrive at a possibly incorrect
assessment of this claim by using the following line of reasoning:

\begin{description}
\item[ \ \ $-$] If there exists a linear map $y-Bu=0$ between $u\in U$ and
$y\in\mathcal{P}_{n}$, then%
\[
\overline{U}:=\left\{  \left(
\begin{array}
[c]{c}%
u\\
y
\end{array}
\right)  \in\mathbb{R}^{\mu+n(n-1)}:Au\leq b;\text{ }y-Bu=0\right\}
\]
is an EF\ of $\mathcal{P}_{n}$, and hence, $U$ cannot be polynomial-sized.

\item[ \ \ $-$] If there does not exist a linear map $y-Bu=0$ between $u\in U$
and $y\in\mathcal{P}_{n}$, then $U$\ cannot be an EF of $\mathcal{P}_{n},$ and
hence, LP cannot solve the TSP.

\item[ \ \ $-$] Hence, one concludes that the claim cannot be true, which may
not be the correct conclusion, as we will develop in this paper.\medskip
\end{description}
\end{claim}

\begin{remark}
[Misconceptions]\label{Misconceptions_Rmk}Some reasons why using the
\textit{extension}-based EF illustrated above in particular in making
inferences about model sizes can lead to overscoped/erroneous conclusions are
as follows:

\begin{enumerate}
\item The lack of distinction between what is/is not a meaningful EF in the
context of making inferences about model sizes;

\item Inferences about solvability for an optimization problem being made
without consideration to the expression of the objective function;

\item The significance of superfluousness (\textquotedblleft
redundancy\textquotedblright) from an optimization problem prespective being overlooked;

\item The implicit presumption that the TSP optimization problem in its
natural form (i.e., without reduction to the optimization version of another
NPC problem) can be solved\ only through \textquotedblleft\textit{the} TSP
polytope\textquotedblright.
\end{enumerate}
\end{remark}

We will show in the remainder of this paper that provided a given polytope
projects to the LAP polytope, $\mathcal{A}_{n}$, and subject to the
possibility of expressing TSP tour costs as linear functions of its
descriptive variables, an LP over that polytope (or an EF of it) could solve
the TSP optimization problem, irrespective of whether it (or any EF of it)
projects to \textquotedblleft\textit{the} TSP polytope,\textquotedblright%
\ $\mathcal{P}_{n}$, or not.

\section{Side-stepping of the \textit{barriers}\label{SideStepping_Section}}

The key to \textquotedblleft escaping from\textquotedblright\ the
\textit{barriers} lies in the judicious exploitation of the one-to-one
correspondence that exists between the TSP tours and the extreme points of the
LAP polytope $\mathcal{A}_{n}$ (and equivalently, the \textit{TSP paths} of
the TSPAG). Because of this one-to-one correspondence, if tour costs can be
correctly modeled by a linear function of the auxiliary variables of an EF of
$\mathcal{A}_{n}$, then the TSP optimization problem can be solved without
recourse to \textquotedblleft\textit{the} TSP polytope,\textquotedblright%
\ using that EF.\ Our developments to show this consist essentially of
examining the \textit{extensions }of the linear assignment problem (LAP)
polytope summarized in Table $1$. \bigskip%

\begin{tabular}
[c]{ll}\hline
\multicolumn{1}{|l|}{} & \multicolumn{1}{l|}{}\\
\multicolumn{1}{|l|}{%
\begin{tabular}
[c]{l}%
\textit{Polytope }$\mathcal{A}_{n}:$%
\end{tabular}
} & \multicolumn{1}{l|}{%
\begin{tabular}
[c]{l}%
\textit{Polytope }$Q_{0}:$%
\end{tabular}
}\\
\multicolumn{1}{|l}{} & \multicolumn{1}{|l|}{}\\
\multicolumn{1}{|l}{%
\begin{tabular}
[c]{ll}
& $\sum\limits_{r=1}^{m}w_{ir}\text{ \ }=\text{ \ }1;\text{ }(\forall
i=1,\ldots,m).$%
\end{tabular}
} & \multicolumn{1}{|l|}{%
\begin{tabular}
[c]{ll}
& $w\in\mathcal{A}_{n}.$%
\end{tabular}
}\\
\multicolumn{1}{|l}{} & \multicolumn{1}{|l|}{}\\
\multicolumn{1}{|l|}{%
\begin{tabular}
[c]{ll}
& $\sum\limits_{i=1}^{m}w_{ir}\text{ \ }=\text{ \ }1;\text{ }(\forall
r=1,\ldots,m).$%
\end{tabular}
} & \multicolumn{1}{l|}{%
\begin{tabular}
[c]{ll}
& $\mathbf{A}w\mathbf{+B}x\mathbf{=b.}$%
\end{tabular}
}\\
\multicolumn{1}{|l}{} & \multicolumn{1}{|l|}{}\\
\multicolumn{1}{|l|}{%
\begin{tabular}
[c]{ll}
& $w_{ir}\text{ \ }\geq\text{\ }0;$ $(\forall i,r=1,\ldots,m).$%
\end{tabular}
} & \multicolumn{1}{l|}{%
\begin{tabular}
[c]{ll}
& ($\mathbf{A}\in\mathbb{R}^{\alpha\times m^{2}};\text{ }\mathbf{B}%
\in\mathbb{R}^{\alpha\times\kappa};\text{ }\mathbf{b}\in\mathbb{R}^{\alpha
}.\text{)}$%
\end{tabular}
}\\
\multicolumn{1}{|l|}{} & \multicolumn{1}{l|}{}\\\hline
\multicolumn{1}{|l|}{} & \multicolumn{1}{l|}{}\\
\multicolumn{1}{|l}{%
\begin{tabular}
[c]{l}%
\textit{Polytope }$Q_{1}:$%
\end{tabular}
} & \multicolumn{1}{|l|}{%
\begin{tabular}
[c]{l}%
\textit{Polytope }$Q_{2}:$%
\end{tabular}
}\\
\multicolumn{1}{|l}{} & \multicolumn{1}{|l|}{}\\
\multicolumn{1}{|l|}{%
\begin{tabular}
[c]{ll}
& $w\in\mathcal{A}_{n}.$%
\end{tabular}
} & \multicolumn{1}{l|}{%
\begin{tabular}
[c]{ll}
& $\left(
\begin{array}
[c]{c}%
w\\
x
\end{array}
\right)  \in Q_{0}.$%
\end{tabular}
}\\
\multicolumn{1}{|l}{} & \multicolumn{1}{|l|}{}\\
\multicolumn{1}{|l|}{%
\begin{tabular}
[c]{ll}
& $\mathbf{E}^{\mathbf{=}}w$ $\mathbf{+}$ $\mathbf{F}^{\mathbf{=}}y$
$\mathbf{=}$ $\mathbf{f}^{\mathbf{=}}.$%
\end{tabular}
} & \multicolumn{1}{l|}{%
\begin{tabular}
[c]{ll}
& $\mathbf{G}^{\mathbf{=}}w$ $\mathbf{+}$ $\mathbf{H}^{\mathbf{=}}y$
$\mathbf{=}$ $\mathbf{h}^{\mathbf{=}}.$%
\end{tabular}
}\\
\multicolumn{1}{|l}{} & \multicolumn{1}{|l|}{}\\
\multicolumn{1}{|l|}{%
\begin{tabular}
[c]{ll}
& $\mathbf{E}^{\mathbf{\leq}}w$ $\mathbf{+}$ $\mathbf{F}^{\mathbf{\leq}}y$
$\mathbf{\leq}$ $\mathbf{f}^{\mathbf{\leq}}.$%
\end{tabular}
} & \multicolumn{1}{l|}{%
\begin{tabular}
[c]{ll}
& $\mathbf{G}^{\mathbf{\leq}}w$ $\mathbf{+}$ $\mathbf{H}^{\mathbf{\leq}}y$
$\mathbf{\leq}$ $\mathbf{h}^{\mathbf{\leq}}.$%
\end{tabular}
}\\
\multicolumn{1}{|l}{} & \multicolumn{1}{|l|}{}\\
\multicolumn{1}{|l|}{%
\begin{tabular}
[c]{ll}
& $\text{(}\mathbf{E}^{\mathbf{=}}\in\mathbb{R}^{\beta^{=}\times m^{2}};\text{
}\mathbf{F}^{\mathbf{=}}\in\mathbb{R}^{\beta^{=}\times\ell};\text{ }$%
\end{tabular}
} & \multicolumn{1}{l|}{%
\begin{tabular}
[c]{ll}
& $\text{(}\mathbf{G}^{\mathbf{=}}\in\mathbb{R}^{\gamma^{=}\times m^{2}%
};\text{ }\mathbf{H}^{\mathbf{=}}\in\mathbb{R}^{\gamma^{=}\times\ell};$%
\end{tabular}
}\\
\multicolumn{1}{|l}{} & \multicolumn{1}{|l|}{}\\
\multicolumn{1}{|l}{%
\begin{tabular}
[c]{ll}
& $\mathbf{f}^{\mathbf{=}}\in\mathbb{R}^{\beta^{=}};$ $\mathbf{E}%
^{\mathbf{\leq}}\in\mathbb{R}^{\beta^{\leq}\times m^{2}};\text{ }$%
\end{tabular}
} & \multicolumn{1}{|l|}{%
\begin{tabular}
[c]{ll}
& $\mathbf{h}^{\mathbf{=}}\in\mathbb{R}^{\gamma^{=}};\text{ }\mathbf{G}%
^{\mathbf{\leq}}\in\mathbb{R}^{\gamma^{\leq}\times m^{2}};$%
\end{tabular}
}\\
\multicolumn{1}{|l}{} & \multicolumn{1}{|l|}{}\\
\multicolumn{1}{|l|}{%
\begin{tabular}
[c]{ll}
& $\mathbf{F}^{\mathbf{\leq}}\in\mathbb{R}^{\beta^{\leq}\times\ell};\text{
}\mathbf{f}^{\mathbf{\leq}}\in\mathbb{R}^{\beta^{\leq}}.\text{)}$%
\end{tabular}
} & \multicolumn{1}{l|}{%
\begin{tabular}
[c]{ll}
& $\mathbf{H}^{\mathbf{\leq}}\in\mathbb{R}^{\gamma^{\leq}\times\ell};\text{
}\mathbf{h}^{\mathbf{\leq}}\in\mathbb{R}^{\gamma^{\leq}}.\text{)}$%
\end{tabular}
}\\
\multicolumn{1}{|l|}{} & \multicolumn{1}{l|}{}\\\hline
& \\
\multicolumn{2}{c}{Table 1: Statements of the Polytopes}\\
&
\end{tabular}

\textit{Polytope} $\mathcal{A}_{n}$ is the LAP polytope explicitly stated in
the space of the $w$ variables. \textit{Polytope} $Q_{0}$ is in the space of
$w$ and its \textquotedblleft auxiliary\textquotedblright\ variables $x.$
\textit{Polytope} $Q_{1}$ is in the space of $w$ and TSP \textit{travel leg}
variables $y$. \textit{Polytope} $Q_{2}$ is in the space of all of the three
classes of variables $w$, $x$ and $y$.

The optimization problems associated with the polytopes described above (Table
$1$) are summarized in Table $2,$ where $c\in\mathbb{R}^{m^{2}}$ and
$\widetilde{c}\in\mathbb{R}^{\ell}$. \textit{Problem }$LAP$ is the
\textquotedblleft standard\textquotedblright\ linear assignment problem (LAP)
stated in the space of $w$. \textit{Problem} $LP_{0}$ seeks to minimize a
linear function of $x$ over $Q_{0}$. \textit{Problem} $LP_{1}$ seeks to
minimize a linear function of $w$ over $Q_{1}$. \textit{Problem} $LP_{2}$ also
seeks to minimize a linear function of $x$, but the applicable constraints set
is $Q_{2}$. \bigskip

\noindent%
\begin{tabular}
[c]{ll}\hline
\multicolumn{1}{|l|}{} & \multicolumn{1}{l|}{}\\
\multicolumn{1}{|l}{\textit{Problem} $LAP:$} &
\multicolumn{1}{|l|}{\textit{Problem}\textbf{\ } $LP_{\mathbf{0}}:$}\\
\multicolumn{1}{|c}{} & \multicolumn{1}{|c|}{}\\
\multicolumn{1}{|l|}{%
\begin{tabular}
[c]{l}%
\ \ \
\end{tabular}
$\left.
\begin{array}
[c]{l}%
\text{Minimize }\text{: \ \ }c^{t}\mathit{w}\text{ \ \ }\\
\text{ \ \ }\\
\text{Subject to}\text{: \ }\mathit{w}\in\mathcal{A}_{n}.
\end{array}
\right.  $} & \multicolumn{1}{l|}{%
\begin{tabular}
[c]{l}%
\ \ \
\end{tabular}
$\left.
\begin{array}
[c]{l}%
\text{Minimize }\text{: \ \ }\widetilde{c}^{t}\mathit{x}\text{ \ \ }\\
\text{ \ \ }\\
\text{Subject to}\text{: \ }\left(
\begin{array}
[c]{c}%
w\\
x
\end{array}
\right)  \in Q_{0}.
\end{array}
\right.  $}\\
\multicolumn{1}{|l|}{} & \multicolumn{1}{l|}{}\\\hline
\multicolumn{1}{|l|}{} & \multicolumn{1}{l|}{}\\
\multicolumn{1}{|l|}{\textit{Problem }$LP_{\mathbf{1}}:$} &
\multicolumn{1}{l|}{\textit{Problem}\textbf{\ }$LP_{\mathbf{2}}:$}\\
\multicolumn{1}{|c}{} & \multicolumn{1}{|c|}{}\\
\multicolumn{1}{|l}{%
\begin{tabular}
[c]{l}%
\ \ \
\end{tabular}
$\left.
\begin{array}
[c]{l}%
\text{Minimize }\text{: \ \ }c^{t}\mathit{w}\text{ \ \ }\\
\text{ \ \ }\\
\text{Subject to}\text{: \ }\left(
\begin{array}
[c]{c}%
w\\
y
\end{array}
\right)  \in Q_{1}.
\end{array}
\right.  $} & \multicolumn{1}{|l|}{%
\begin{tabular}
[c]{l}%
\ \ \
\end{tabular}
$\left.
\begin{array}
[c]{l}%
\text{Minimize }\text{: \ \ }\widetilde{c}^{t}\mathit{x}\text{ \ \ }\\
\text{ \ \ }\\
\text{Subject to}\text{: \ }\left(
\begin{array}
[c]{c}%
w\\
x\\
y
\end{array}
\right)  \in Q_{2}.
\end{array}
\right.  $}\\
\multicolumn{1}{|l|}{} & \multicolumn{1}{l|}{}\\\hline
& \\
\multicolumn{2}{c}{Table 2: Statements of the Optimization Problems}\\
&
\end{tabular}

Observe that because costs are not put on the \textit{travel leg} variables,
$y$, \textit{if} \textit{Polytope} $Q_{1}$ is assumed to be an EF of
\textit{Polytope }$\mathcal{A}_{n}$, \textit{then} \textit{Problem }$LAP$ and
\textit{Problem }$LP_{1}$ would be equivalent optimization problems in the
sense that they would have the same set of optimal $w$-variables and the same
objective value. Similarly, since costs are not put on the \textit{travel leg}
variables, $y$, in \textit{Problem }$LP_{2},$ \textit{Problem }$LP_{2}$ and
\textit{Problem }$LP_{0}$ would be equivalent optimization problems
\textit{if} $Q_{2}$ is an EF of $Q_{0}$.

The equivalence between \textit{Problem }$LP_{2}$ and \textit{Problem }%
$LP_{0}$\ is the one that is essential in our argumentation. The equivalence
between \textit{Problem }$LAP$ and \textit{Problem }$LP_{1}$, however, is
particularly easy to use in developing an illustration, as we will do next.

\subsection{A motivating illustrative
example\label{Illutrative_Example_SubSection}\medskip}

The example we discuss has the general form of \textit{Problem}
$LP_{\mathbf{1}}$. The polytope involved is as follows: \ %

\begin{align}
\mathit{Polytope\ }\overline{Q}_{1}: &  \nonumber\\[0.06in]
&  \sum\limits_{r=1}^{m}w_{ir}\text{ \ }=\text{ \ }1;\text{ }(\forall
i=1,\ldots,m).\text{ }\label{Example(1)}\\[0.03in]
&  \sum\limits_{i=1}^{m}w_{ir}\text{ \ }=\text{ \ }1;\text{ }(\forall
r=1,\ldots,m).\label{Example(2)}\\[0.09in]
&  w_{ir}\text{ }\geq\text{\ }0;\text{ }(\forall i,r=1,\ldots
,m).\label{Example(3)}\\[0.15in]
&  y_{0,i}-w_{i,1}=0;\text{ }(\forall i=1,\ldots,m).\label{Example(4)}%
\\[0.15in]
&  y_{i,0}-w_{i,m}=0;\text{\ }(\forall i=1,\ldots,m).\label{Example(5)}%
\\[0.12in]
&  w_{ir}+w_{j,r+1}-y_{ij}\leq\text{ \ }1;\text{ }(\forall i,r,j=1,\ldots
,m\text{ }(r\leq m-1\text{ }\wedge\text{\ }i\neq j)).\text{ }%
\label{Example(6)}\\[0.06in]
&  \sum\limits_{i=1}^{m}\sum\limits_{j=1;j\neq i}^{m}y_{ij}=\text{
\ }m-1.\label{Example(7)}\\[0.06in]
&  y_{ij}\geq\text{\ }0;\text{ }(\forall i,j=0,\ldots,m).\label{Example(8)}%
\end{align}

Constraints $(\ref{Example(1)})-(\ref{Example(3)})$ are those of
\textit{Polytope} $\mathcal{A}_{n}$. They are stated again for the sake of
completeness and convenience. The optimization problem associated with this
polytope is \textit{Problem} $LAP,$ as shown in Table $2$. Constraints
$(\ref{Example(4)})-(\ref{Example(8)})$ are a special case of the
additional/extension constraints of \textit{Polytope} $Q_{1}$. We will denote
this special case of $Q_{1}$ (i.e., constraints $(\ref{Example(1)}%
)-(\ref{Example(8)})$) and its associated optimization problem by
$\overline{Q}_{1}$ and $\overline{LP}_{1}$, respectively. In other words,
\textit{Problem }$\overline{LP}_{1}$ is as follows:

\ %

\begin{tabular}
[c]{l}%
\textit{Problem }$\overline{LP}_{1}$:\\
\ \ \\%
\begin{tabular}
[c]{l}%
\ \ \
\end{tabular}
$\left\vert
\begin{array}
[c]{l}%
\text{Minimize }\text{: \ \ }\sum\limits_{i=1}^{m}\sum\limits_{r=1}^{m}%
c_{ir}w_{ir}\text{ \ \ }\\
\text{ \ \ }\\
\text{Subject to}\text{: }(\ref{Example(1)})-(\ref{Example(8)}).
\end{array}
\right.  $\\
\ \ \ \ \
\end{tabular}
\ \  \bigskip

\ 

We have the following results.

\begin{lemma}
\label{Example_Lemma_1} $\overline{Q}_{1}$ is an EF of $\mathcal{A}_{n}.$
\end{lemma}

\begin{proof}
We want to show that%
\[
\pi_{w}(\overline{Q}_{1}):=\left\{  w\in\mathbb{R}^{m^{2}}:\left(  \exists
y\in\mathbb{R}^{nm}:\left(
\begin{array}
[c]{c}%
w\\
x
\end{array}
\right)  \in\overline{Q}_{1}\right)  \right\}  =\mathcal{A}_{n}.
\]

\begin{enumerate}
\item[$(a)$] Since the descriptive constraints of $\mathcal{A}_{n}$
$((\ref{Example(1)})$-$(\ref{Example(3)}))$ are included in the description of
$\overline{Q}_{1}$,
\begin{equation}
\left(
\begin{array}
[c]{c}%
w\\
x
\end{array}
\right)  \in\overline{Q}_{1}\implies w\in\mathcal{A}_{n}. \label{Expl_Lem1(1)}%
\end{equation}

$(\ref{Expl_Lem1(1)})$ implies:%
\begin{equation}
\pi_{w}(\overline{Q}_{1})\subseteq\mathcal{A}_{n}. \label{Expl_Lem1(2)}%
\end{equation}

\item[$(b)$] Let $\widehat{w}\in Ext(\mathcal{A}_{n}).$ Let $\widehat{y}%
\in\mathbb{R}^{nm}$ be specified in terms of $\widehat{w}$ as follows:%
\begin{equation}
\forall i\in\{1,\ldots,m\},\text{ }\widehat{y}_{0,i}=\left\{
\begin{array}
[c]{l}%
1\text{ \ \ if }\widehat{w}_{i,1}=1\\
\\
0\text{ \ \ otherwise}%
\end{array}
\right.  \label{Expl_Lem1(3)}%
\end{equation}%
\begin{equation}
\forall i\in\{1,\ldots,m\},\text{ }\widehat{y}_{i,0}=\left\{
\begin{array}
[c]{l}%
1\text{ \ \ if }\widehat{w}_{i,m}=1\\
\\
0\text{ \ \ otherwise}%
\end{array}
\right.  \label{Expl_Lem1(4)}%
\end{equation}%
\begin{equation}
\forall i,j\in\{1,\ldots,m\},\text{ }\widehat{y}_{ij}=\left\{
\begin{array}
[c]{l}%
1\text{ \ \ if }\exists r\in\{1,\ldots,m-1\}:\text{ }\widehat{w}%
_{ir}=\widehat{w}_{j,r+1}=1.\\
\\
0\text{ \ \ otherwise}%
\end{array}
\right.  \label{Expl_Lem1(5)}%
\end{equation}

Observe that $(\ref{Expl_Lem1(3)})$-$(\ref{Expl_Lem1(5)}),$ $(\ref{Example(1)}%
)$-$(\ref{Example(3)})$, and the integrality of $\widehat{w}$ imply:%
\begin{align}
\exists!i  &  \in C:\widehat{y}_{0,i}=1;\label{Expl_Lem1(6)}\\[0.09in]
\exists!i  &  \in C:\widehat{y}_{i,0}=1;\label{Expl_Lem1(7)}\\[0.09in]
\forall r  &  \in\{1,\ldots,m-1\},\exists!(i,j)\in C^{2}:\widehat{y}_{ij}=1.
\label{Expl_Lem1(8)}%
\end{align}
Observe also that these $y$-variables (i.e., as set according to
$(\ref{Expl_Lem1(3)})$-$(\ref{Expl_Lem1(5)})$) also satisfy $(\ref{Example(7)}%
).$ Hence, $\left(
\begin{array}
[c]{c}%
\widehat{w}\\
\widehat{y}%
\end{array}
\right)  $ satisfies $(\ref{Example(1)})$-$(\ref{Example(8)}).$ In other
words, we have:
\begin{equation}
\left(
\begin{array}
[c]{c}%
\widehat{w}\\
\widehat{y}%
\end{array}
\right)  \in\overline{Q}_{1}. \label{Expl_Lem1(9)}%
\end{equation}

($\ref{Expl_Lem1(9)}$) and the arbitrariness of $\widehat{w}$ imply:%
\begin{equation}
\widehat{w}\in Ext(\mathcal{A}_{n})\implies\exists\widehat{y}\in
\mathbb{R}^{nm}:\left(
\begin{array}
[c]{c}%
\widehat{w}\\
\widehat{x}%
\end{array}
\right)  \in\overline{Q}_{1}. \label{Expl_Lem1(10)}%
\end{equation}
($\ref{Expl_Lem1(10)}$) implies:%
\begin{equation}
Conv(Ext(\mathcal{A}_{n}))=\mathcal{A}_{n}\subseteq\pi_{w}(\overline{Q}_{1}).
\label{Expl_Lem1(11)}%
\end{equation}

\item[$(c)$] ($\ref{Expl_Lem1(2)}$) and ($\ref{Expl_Lem1(11)}$) imply:%
\[
\pi_{w}(\overline{Q}_{1})=\mathcal{A}_{n}.
\]

\end{enumerate}
\end{proof}

\begin{lemma}
\label{Example_Lemma_2}$\overline{Q}_{1}$ is not an EF of\ $\mathcal{P}_{n}$.
Moreover, $\overline{Q}_{1}$ is non-integral.
\end{lemma}

\begin{proof}
First, we will show that:
\begin{equation}
\left(
\begin{array}
[c]{c}%
w\\
y
\end{array}
\right)  \in\overline{Q}_{1}\nRightarrow y\in\mathcal{P}_{n}.
\label{Expl_Lem2(0)}%
\end{equation}
The proof is by construction. Consider the special case of $\overline{Q}_{1}$
for a $5$-city TSP. \ Consider $\left(
\begin{array}
[c]{c}%
\overline{w}\\
\overline{y}%
\end{array}
\right)  \in\mathbb{R}_{\geq0}^{m^{2}+nm}$ which has the following non-zero
components:%
\begin{equation}
\left\{
\begin{array}
[c]{l}%
\overline{w}_{1,1}=\overline{w}_{2,2}=\overline{w}_{4,3}=\overline{w}%
_{3,4}=0.25;\\
\\
\overline{w}_{4,1}=\overline{w}_{3,2}=\overline{w}_{1,3}=\overline{w}%
_{2,4}=0.75;\\
\\
\overline{y}_{0,1}=\overline{y}_{3,0}=0.25;\text{ }\overline{y}_{0,4}%
=\overline{y}_{2,0}=0.75;\\
\\
\overline{y}_{3,1}=\overline{y}_{1,2}=0.50;\text{ }\overline{y}_{4,3}=2.0.
\end{array}
\right.  \text{ } \label{Expl_Lem2(0.1)}%
\end{equation}

Observe that $\left(
\begin{array}
[c]{c}%
\overline{w}\\
\overline{y}%
\end{array}
\right)  $ satisfies $(\ref{Example(1)})-(\ref{Example(8)})$ for $n=5$. Also,
since every point of \textquotedblleft\textit{the} TSP
polytope\textquotedblright\ has coordinates between $0$ and $1,$ $\overline
{y}_{4,3}$ being greater than $1$ ($\overline{y}_{4,3}=2>1$),\ in particular,
implies that $\overline{y}\notin\mathcal{P}_{5}.$ Statement
($\ref{Expl_Lem2(0)}$) follows directly from these observations.

Focusing back on $\mathcal{P}_{n}$\ and $\overline{Q}_{1}$ in general, observe
that every feasible solution to $(\ref{Example(1)})-(\ref{Example(8)})$ is
such that its $y$-vector component is integral if and only if its $w$-vector
component is integral. In other words, it can be observed that the following
is true:
\begin{equation}
\left(
\begin{array}
[c]{c}%
w\\
y
\end{array}
\right)  \in\overline{Q}_{1}\implies\left(  w\in\{0,1\}^{m^{2}}\iff
y\in\{0,1\}^{nm}\right)  . \label{Expl_Lem2(1)}%
\end{equation}
The combination of the fact that every extreme point of $\mathcal{A}_{n}$ is
an integral $w$-vector, statement $(\ref{Expl_Lem2(1)})$, and Remark
\ref{Path&Tour_Correspondence_Remark} gives that:%
\begin{align}
&  \left(  \left\{  y\in\{0,1\}^{nm}:\left(  \exists w\in\{0,1\}^{m^{2}%
}:\left(
\begin{array}
[c]{c}%
w\\
y
\end{array}
\right)  \in\overline{Q}_{1}\right)  \right\}  \right)  \text{ is
}\nonumber\\[0.06in]
&  \text{in a one-to-one correspondence with the set of TSP tours.}
\label{Expl_Lem2(2)}%
\end{align}
($\ref{Expl_Lem2(2)}$) implies:%
\begin{equation}
\mathcal{P}_{n}=Conv\left\{  y\in\{0,1\}^{nm}:\left(  \exists w\in
\{0,1\}^{m^{2}}:\left(
\begin{array}
[c]{c}%
w\\
y
\end{array}
\right)  \in\overline{Q}_{1}\right)  \right\}  \text{.} \label{Expl_Lem2(2.2)}%
\end{equation}
($\ref{Expl_Lem2(2.2)}$) implies:
\begin{equation}
\mathcal{P}_{n}\subseteq\left\{  y\in\mathbb{R}^{nm}:\left(  \exists
w\in\mathbb{R}^{m^{2}}:\left(
\begin{array}
[c]{c}%
w\\
y
\end{array}
\right)  \in\overline{Q}_{1}\right)  \right\}  . \label{Expl_Lem2(2.5)}%
\end{equation}
From ($\ref{Expl_Lem2(0)}$), we have:%
\begin{equation}
\left\{  y\notin\mathcal{P}_{n}:\left(  \exists w\in\mathbb{R}^{m^{2}}:\left(
\begin{array}
[c]{c}%
w\\
y
\end{array}
\right)  \in\overline{Q}_{1}\right)  \right\}  \neq\varnothing.
\label{Expl_Lem2(3)}%
\end{equation}
($\ref{Expl_Lem2(2.5)}$) and ($\ref{Expl_Lem2(3)}$) imply:%
\begin{equation}
\mathcal{P}_{n}\subset\left\{  y\in\mathbb{R}^{mn}:\left(  \exists
w\in\mathbb{R}^{m^{2}}:\left(
\begin{array}
[c]{c}%
w\\
y
\end{array}
\right)  \in\overline{Q}_{1}\right)  \right\}  . \label{Expl_Lem2(3.5)}%
\end{equation}
Hence,
\begin{equation}
\left\{  y\in\mathbb{R}^{mn}:\left(  \exists w\in\mathbb{R}^{m^{2}}:\left(
\begin{array}
[c]{c}%
w\\
y
\end{array}
\right)  \in\overline{Q}_{1}\right)  \right\}  \neq\mathcal{P}_{n}.
\label{Expl_Lem2(3.7)}%
\end{equation}
Hence, $\overline{Q}_{1}$ is not an EF of $\mathcal{P}_{n}.$

Finally (thirdly), we will show that $\overline{Q}_{1}$ is non-integral. To
that effect, let $I(\overline{Q}_{1})$ be the set of integral solutions of
$\overline{Q}_{1}.$ Then, we have the following.

Statements ($\ref{Expl_Lem2(2.2)}$) and ($\ref{Expl_Lem2(3.5)}$) imply:
\begin{equation}
\overline{Q}_{1}\backslash Conv(I(\overline{Q}_{1}))\neq\varnothing.
\label{Expl_Lem2(5)}%
\end{equation}

Since every feasible point of a polytope must be a convex combination of
extreme points of that polytope (Bazaraa et al. [2010; pp. 70-82]), and since
$\left(
\begin{array}
[c]{c}%
\widetilde{w}\\
\widetilde{y}%
\end{array}
\right)  \in\overline{Q}_{1}\backslash Conv(I(\overline{Q}_{1}))$ cannot be a
convex combination of integral points of $\overline{Q}_{1}$ (since $\left(
\begin{array}
[c]{c}%
\widetilde{w}\\
\widetilde{y}%
\end{array}
\right)  \notin Conv(I(\overline{Q}_{1}))$), at least one of the extreme
points of $\overline{Q}_{1}$\ with positive weights in the representation of
$\left(
\begin{array}
[c]{c}%
\widetilde{w}\\
\widetilde{y}%
\end{array}
\right)  $ must be non-integral. Hence, $\overline{Q}_{1}$ must have at least
one non-integral extreme point. Hence, $\overline{Q}_{1}$ is non-integral.
\ \ \medskip
\end{proof}

The following result is proved more generally in the next section for
\textit{Problem} $LAP$ and \textit{Problem} $LP_{1}$ (Theorem
$\ref{LP_Equivalences-1}$ of section $\ref{Gen'l_Result_Section}$ ). Hence,
its proof is omitted in this section.

\begin{lemma}
\label{Example_Lemma_3}$\left(
\begin{array}
[c]{c}%
w^{\ast}\\
y^{\ast}%
\end{array}
\right)  $ is optimal for Problem $\overline{LP}_{1}$ iff $w^{\ast}$ is
optimal for \textit{Problem} $\overline{LAP}$. \ 
\end{lemma}

The main take-aways from this section are summarized\ in the following remark. \ \ 

\begin{remark}
\label{Example1_Rmk} \ \ 

\begin{enumerate}
\item Polytope $\overline{Q}_{1}$ is an EF of the LAP polytope, $\mathcal{A}%
_{n}$.

\item Problem $\overline{LP}_{1}$ and Problem $LAP$ have the same objective
value and the same set of optimal $w$ variables.

\item If the optimum in $(2)$ above is unique, it is integral and must be
obtained whether Problem $LAP$ or Problem $\overline{LP}_{1}$ is used/solved.

\item If there are multiple optima to Problem $\overline{LP}_{1}$ (or
equivalently to Problem $LAP$), then:

\begin{enumerate}
\item There must exist more than one integral extreme point of $\overline
{Q}_{1}$ ($\mathcal{A}_{n}$) that is optimal for \textit{Problem} $LP_{1}$
(\textit{Problem} $LAP$);

\item A fractional solution which is a convex combination of the optimal
extreme points in ($a$) above for $\overline{Q}_{1}$ (i.e., an interior point
of$\mathcal{\ }\overline{Q}_{1}$) may be obtained when \textit{Problem}
$LP_{1}$ is solved.

\item The $w$-component of any fractional solution of $\overline{Q}_{1}$
obtained as described in $(b)$ above would be an interior point of
$\mathcal{A}_{n}$, and would hence be a convex combination of integral $w$'s.
Each of these integral $w$'s would correspond to exactly one TSP tour.
\end{enumerate}

\item Statements ($1$)-($4$) of Remark \ref{Example1_Rmk}\ above are true,
irrespective of whether an EF of \textit{Polytope} $\overline{Q}_{1}$ projects
to \textquotedblleft\textit{the} TSP polytope,\textquotedblright%
\ $\mathcal{P}_{n}$, or not.\ \ 
\end{enumerate}
\end{remark}

The following example illustrates the fact that an LP with feasible set
$\overline{Q}_{1}$ (or an EF of it) could solve the TSP optimization problem
despite the fact that $\overline{Q}_{1}$ is not integral and \textit{does not
project} to \textquotedblleft\textit{the} TSP polytope,\textquotedblright%
\ $\mathcal{P}_{n}$. This exemplifies the fact that an LP could solve the TSP
optimization problem, irrespective of whether it projects to $\mathcal{P}_{n}$
or not, provided it project to $\mathcal{A}_{n}.$

\begin{example}
\label{Illustr-Expl_Example}Re-consider the special case of $\overline{Q}_{1}$
for a $5$-city TSP. Consider the solution instance shown in
(\ref{Expl_Lem2(0.1)}). We will show that this solution can be associated with
TSP\ tours by focusing on $\overline{w}$. To this effect, let $\widehat{w}%
^{1}$ and $\widehat{w}^{2}$ be extreme points of $\mathcal{A}_{5}$ with
positive components as follows:%
\[
\left\{
\begin{array}
[c]{l}%
\widehat{w}_{1,1}^{1}=\widehat{w}_{2,2}^{1}=\widehat{w}_{4,3}^{1}%
=\widehat{w}_{3,4}^{1}=1;\\
\\
\widehat{w}_{4,1}^{2}=\widehat{w}_{3,2}^{2}=\widehat{w}_{1,3}^{2}%
=\widehat{w}_{2,4}^{2}=1.
\end{array}
\right.  \text{ }%
\]

Then, by the one-to-one correspondence between TSP tours and the extreme
points of $\mathcal{A}_{5},$ there are unique TSP tours associated with
$\widehat{w}^{1}$ and $\widehat{w}^{2},$ respectively. The tour represented by
$\widehat{w}^{1},$ $\tau(\widehat{w}^{1}),$ is: $0\longrightarrow
1\longrightarrow2\longrightarrow4\longrightarrow3\longrightarrow0.$ Similarly,
the tour, $\tau(\widehat{w}^{2}),$ represented by $\widehat{w}^{2} $ is:
$0\longrightarrow4\longrightarrow3\longrightarrow1\longrightarrow
2\longrightarrow0.$

Observe that $\overline{w}=0.25\widehat{w}^{1}+0.75\widehat{w}^{2}$ is a
convex combination of $\widehat{w}^{1}$ and $\widehat{w}^{2}.$ Hence, $\left(
\begin{array}
[c]{c}%
\overline{w}\\
\overline{y}%
\end{array}
\right)  $ would be associated with $\tau(\widehat{w}^{1})$ and $\tau
(\widehat{w}^{2})$ and their convex combination in any optimization problem
over $\overline{Q}_{1}$ in which the objective function is a linear function
of $w$.

Observe also that the associations above for $\left(
\begin{array}
[c]{c}%
\overline{w}\\
\overline{y}%
\end{array}
\right)  $ analogously hold for every $\left(
\begin{array}
[c]{c}%
w\\
y
\end{array}
\right)  \in\overline{Q}_{1}$. In other words, every $\left(
\begin{array}
[c]{c}%
w\\
y
\end{array}
\right)  \in\overline{Q}_{1}$ would be associated to a number $\eta
\in\mathbb{N}\ $of TSP tours $\tau_{1}(w),\ldots,\tau_{\eta}(w)$ and their
convex combination in any optimization problem over $\overline{Q}_{1}$ in
which the objective function is a linear function of $w$. Hence, if costs can
be attached to $w$ through a linear function in such a way that the total cost
incurred at an extreme point of $\mathcal{A}_{n},$ $\widehat{w}^{i}$
$(i\in\{1,\ldots,m!\}),$ is equal to the total cost of the travels involved in
the associated TSP tour, $\tau(\widehat{w}^{i})$, then the TSP optimization
problem would be solved as an LP over $\overline{Q}.$ We believe it is not
possible to \textquotedblleft construct\textquotedblright\ such a cost
association using $w$ directly. However, as we will illustrate in the next
section (Theorem $\ref{Applied_Costs_Thm}$), it is possible to
\textquotedblleft construct\textquotedblright\ such a cost association
indirectly, using auxiliary variables which are based on $w$ only. A
\textit{sufficient} condition for the model which includes these auxiliary
variables (whatever they may end up being) to be able to solve the TSP
optimization problem is that it (the model) be an EF of $\mathcal{A}_{n}$.
Observe that, in other words, whether such an EF of $\overline{Q}_{1}$
projects to \textquotedblleft\textit{the} TSP polytope\textquotedblright%
\ ($\mathcal{P}_{n}$) or not is irrelevant to its being able to solve the TSP
optimization problem. \ \ 
\end{example}

\subsection{General results\label{Gen'l_Result_Section}\ \ }

In this section, we will formally develop our thesis that a finding that there
exists no polynomial-sized integral extended formulation of \textquotedblleft%
\textit{the} TSP polytope,\textquotedblright\ $\mathcal{P}_{n}$, does not
imply that it is \textquotedblleft impossible\textquotedblright\ for a
polynomial-sized linear program to solve the TSP optimization problem. The
general \textquotedblleft flow\textquotedblright\ of our argumentation is as follows:

\begin{description}
\item[$(a)$] \ \ The extreme points of the LAP polytope, $\mathcal{A}_{n}$,
are in one-to-one correspondence with the TSP tours.

\item[$(b)$] \ \ \textit{Provided} the descriptive variables for
$\mathcal{P}_{n}$,\ $y,$ and the constraints involving them in \textit{Problem
}$LP_{2}$ can be discarded from \textit{Problem }$LP_{2}$ (hence reducing
\textit{Problem }$LP_{2}$ to \textit{Problem }$LP_{0}$) without changing the
optimization problem at hand, \textit{Problem }$LP_{2}$ is an equivalent
optimization problem to \textit{Problem }$LP_{0}.$ This equivalence is true
regardless of whether \textit{Polytope} $Q_{1}$ projects to \textit{Polytope}
$\mathcal{P}_{n}$\ or not.

(In the sense above, the $y$-variables and the constraints which involve them
in \textit{Problem }$LP_{2}$ would be \textquotedblleft
redundant\textquotedblright\ for \textit{Problem }$LP_{2}$ (and equivalently
for \textit{Problem} $LP_{0}$), and they could be thought of as only serving
\textquotedblleft bookkeeping\textquotedblright\ purposes in \textit{Problem
}$LP_{2}$. Also, a sufficient condition for this \textit{redundancy} is that
$Q_{2}$ be an EF of $Q_{0}$.)

\item[$(c)$] \ \ The combination of $(a)$ and $(b)$ above implies that
\textit{Problem }$LP_{0}$ would solve the TSP optimization problem,
\textit{provided} the following conditions are satisfied:

\begin{description}
\item[$(i)$] The extreme points of \textit{Polytope} $Q_{0}$ are in one-to-one
correspondence with the extreme points of \textit{Polytope} $\mathcal{A}_{n}$;

\item[$(ii)$] The vector $\widetilde{c}$ is such that it \textquotedblleft
attaches\textquotedblright\ to any given extreme point of $Q_{0}$, the cost of
the TSP tour associated with that extreme point (i.e., for each $\widehat{x}%
\in Ext(Q_{0}),$ $\widetilde{c}^{t}\widehat{x}$ is equal to the cost of the
TSP tour associated with $\widehat{x}$). \ \ 
\end{description}

\item[$(d)$] \ \ Hence, it is true that \textit{Problem }$LP_{0}$ would solve
the TSP optimization problem, regardless of whether \textit{Polytope} $Q_{1}$
projects to \textquotedblleft\textit{the} TSP polytope,\textquotedblright%
\ $\mathcal{P}_{n}$, or not, \textit{provided} the conditions specified in
$(b)$ and $(c)$ above are satisfied.

\item[$(e)$] \ Hence, \textit{provided} the conditions specified in $(b)$ and
$(c)$ above are satisfied, \textquotedblleft\textit{the} TSP
polytope\textquotedblright\ $\mathcal{P}_{n}$ (and hence, the
\textit{barriers}) can be side-stepped via \textit{Problem }$LP_{0}$ in the
process of solving the TSP optimization problem.
\end{description}

These ideas will now be developed formally. We start with the equivalence
between \textit{Problem }$LAP$ and Problem $LP_{1}$ illustrated in section
\ref{Illutrative_Example_SubSection}, although this is not essential to our
argumentation. We will henceforth assume the following extension relationships.

\begin{assumption}
\label{Projection_Assumptions}We will henceforth assume that:

\begin{enumerate}
\item $\pi_{w}(Q_{0})=\left\{  w\in\mathbb{R}^{m^{2}}:\left(  \exists
x\in\mathbb{R}^{k}:\left(
\begin{array}
[c]{c}%
w\\
x
\end{array}
\right)  \in Q_{0}\right)  \right\}  =\mathcal{A}_{n}.$

($Q_{0}$ is an extended formulation (EF) of $\mathcal{A}_{n}$.)

\item $\pi_{w}(Q_{1})=\left\{  w\in\mathbb{R}^{m^{2}}:\left(  \exists
y\in\mathbb{R}^{\ell}:\left(
\begin{array}
[c]{c}%
w\\
y
\end{array}
\right)  \in Q_{1}\right)  \right\}  =\mathcal{A}_{n}.$

($Q_{1}$ is an EF of $\mathcal{A}_{n}$.)

\item $\pi_{w}(Q_{2})=\left\{  w\in\mathbb{R}^{m^{2}}:\left(  \exists\left(
\begin{array}
[c]{c}%
x\\
y
\end{array}
\right)  \in\mathbb{R}^{k+\ell}:\left(
\begin{array}
[c]{c}%
w\\
x\\
y
\end{array}
\right)  (w,x,y)\in Q_{2}\right)  \right\}  =\mathcal{A}_{n}.$

($Q_{2}$ is an EF of $\mathcal{A}_{n}$.)

\item $\pi_{(w,x)}(Q_{2})=\left\{  \left(
\begin{array}
[c]{c}%
w\\
x
\end{array}
\right)  \in\mathbb{R}^{m^{2}+k}:\left(  \exists y\in\mathbb{R}^{\ell}:\left(
\begin{array}
[c]{c}%
w\\
x\\
y
\end{array}
\right)  \in Q_{2}\right)  \right\}  =Q_{0}.$

($Q_{2}$ is an EF of $Q_{0}$.) \ \ \ \medskip
\end{enumerate}
\end{assumption}

Also, the following notations will simplify the exposition. \ \medskip

\begin{notation}
\label{Feasible_Set_Notations_for_w}: \ 

\begin{enumerate}
\item $\forall w\in\mathcal{A}_{n},$ $F_{0}(w):=\left\{  x\in\mathbb{R}%
^{k}:\left(
\begin{array}
[c]{c}%
w\\
x
\end{array}
\right)  \in Q_{0}\right\}  .$\ 

\item $\forall w\in\mathcal{A}_{n},$ $F_{1}(w):=\left\{  y\in\mathbb{R}^{\ell
}:\left(
\begin{array}
[c]{c}%
w\\
y
\end{array}
\right)  \in Q_{1}\right\}  .$

\item $\forall w\in\mathcal{A}_{n},$ $\forall x\in\mathbb{R}^{k}:\left(
\begin{array}
[c]{c}%
w\\
y
\end{array}
\right)  \in Q_{0},$ $F_{2}(w,x):=\left\{  y\in\mathbb{R}^{\ell}:\left(
\begin{array}
[c]{c}%
w\\
x\\
y
\end{array}
\right)  \in Q_{2}\right\}  .$
\end{enumerate}
\end{notation}

The next theorem shows the equivalence between \textit{Problem} $LAP$ and
\textit{Problem }$LP_{1}$, a special case of which was illustrated in section
\ref{Illutrative_Example_SubSection}. \medskip

\begin{theorem}
\label{LP_Equivalences-1}$w^{\ast}$ $\in\mathcal{A}_{n}$ is optimal for
\textit{Problem} $LAP$ if and only if $\left(
\begin{array}
[c]{c}%
w^{\ast}\\
y^{\ast}%
\end{array}
\right)  $ is optimal for \textit{Problem }$LP_{1}$ for each $y^{\ast}\in
F_{1}(w^{\ast}).$
\end{theorem}

\begin{proof}
\ \ \medskip\medskip

$(\implies):$ \ Let $w^{\ast}$ be an optimal solution for \textit{Problem}
$LAP.$ In other words, we have:%
\begin{equation}
\forall w^{\dag}\in\mathcal{A}_{n},\text{ }c^{t}w^{\dag}\geq c^{t}w^{\ast}.
\label{Thm_Eq1(1)}%
\end{equation}
Now, let $y^{\ast}\in F_{1}(w^{\ast}).$ Assume $\left(
\begin{array}
[c]{c}%
w^{\ast}\\
y^{\ast}%
\end{array}
\right)  $\ is not optimal for $Q_{1}.$ Then, there must exist a feasible
solution to \textit{Problem} $LP_{1}$ which has a lower value of the objective
of \textit{Problem} $LP_{1}$ than that of $\left(
\begin{array}
[c]{c}%
w^{\ast}\\
y^{\ast}%
\end{array}
\right)  .$ In other words, we must have:
\begin{equation}
\exists\left(
\begin{array}
[c]{c}%
w^{\dagger}\\
y^{\dagger}%
\end{array}
\right)  \in Q_{1}:\left(  c^{t}w^{\dag}+\mathbf{0}^{t}\cdot y^{\dag}\right)
=c^{t}w^{\dag}<\left(  c^{t}w^{\ast}+\mathbf{0}^{t}\cdot y^{\ast}\right)
=c^{t}w^{\ast}. \label{Thm_Eq1(2)}%
\end{equation}
By Assumption $\ref{Projection_Assumptions}.2$,
\begin{equation}
\left(
\begin{array}
[c]{c}%
w^{\dagger}\\
y^{\dagger}%
\end{array}
\right)  \in Q_{1}\implies w^{\dag}\in\mathcal{A}_{n}. \label{Thm_Eq1(3)}%
\end{equation}
Hence, (\ref{Thm_Eq1(2)}) contradicts (\ref{Thm_Eq1(1)}). Hence, we must have:%
\[
w^{\ast}\text{ optimal for \textit{Problem} }LAP\implies\left(
\begin{array}
[c]{c}%
w^{\ast}\\
y^{\ast}%
\end{array}
\right)  \text{ optimal for }LP_{1}\text{ for each }y^{\ast}\in F_{1}(w^{\ast
}).
\]

$(\impliedby)$: Assume $\left(
\begin{array}
[c]{c}%
w^{\ast}\\
y^{\ast}%
\end{array}
\right)  \in Q_{1}$ is optimal for \textit{Problem} $LP_{1}$ for each
$y^{\ast}\in F_{1}(w^{\ast}).$ In other words, assume the following is true:%
\begin{equation}
\forall\left(
\begin{array}
[c]{c}%
w^{\dagger}\\
y^{\dagger}%
\end{array}
\right)  \in Q_{1},\text{ }\left(  c^{t}w^{\dag}+\mathbf{0}^{t}\cdot y^{\dag
}\right)  =c^{t}w^{\dag}\geq\left(  c^{t}w^{\ast}+\mathbf{0}^{t}\cdot y^{\ast
}\right)  =c^{t}w^{\ast}. \label{Thm_Eq1(5)}%
\end{equation}
Now, assume $w^{\ast}$ is not optimal for \textit{Problem} $LP_{1}$. Then, we
must have:
\begin{equation}
\forall y^{\ast}\in F_{1}(w^{\ast}),\text{ }\exists w^{\dag}\in\mathcal{A}%
_{n}:c^{t}w^{\dag}<c^{t}w^{\ast}=\left(  c^{t}w^{\ast}+\mathbf{0}^{t}\cdot
y^{\ast}\right)  . \label{Thm_Eq1(6)}%
\end{equation}
By Assumption $\ref{Projection_Assumptions}.2$,
\begin{equation}
\left(
\begin{array}
[c]{c}%
w^{\ast}\\
y^{\ast}%
\end{array}
\right)  \in Q_{1}\implies w^{\ast}\in\mathcal{A}_{n}. \label{Thm_Eq1(4)}%
\end{equation}
Hence, statement (\ref{Thm_Eq1(6)}) contradicts statement (\ref{Thm_Eq1(5)}).
Hence, we must have:%
\begin{align}
&  \forall w^{\ast}\in\mathcal{A}_{n},\text{ }\forall y^{\ast}\in
F_{1}(w^{\ast}),\text{ }\left(
\begin{array}
[c]{c}%
w^{\ast}\\
y^{\ast}%
\end{array}
\right)  \text{ optimal for \textit{Problem} }LP_{1}\implies
\nonumber\\[0.06in]
&  w^{\ast}\text{ optimal for \textit{Problem} }LAP\text{. }
\label{Thm_Eq1(4.5)}%
\end{align}
\medskip
\end{proof}

Similar to the relation of \textit{Problem }$LP_{1}$ to \textit{Problem} $LAP
$, \textit{Problem }$LP_{2}$ is equivalent to \textit{Problem} $LP_{0}.$ This
is the object of the next theorem. \ 

\begin{theorem}
\label{LP_Equivalences-2}$\left(
\begin{array}
[c]{c}%
w^{\ast}\\
x^{\ast}%
\end{array}
\right)  \in Q_{0}$ is optimal for \textit{Problem} $LP_{0}$ if and only if
$\left(
\begin{array}
[c]{c}%
w^{\ast}\\
x^{\ast}\\
y^{\ast}%
\end{array}
\right)  $ is optimal for \textit{Problem} $LP_{2}$ for each $y^{\ast}\in
F_{2}(w^{\ast},x^{\ast}).$
\end{theorem}

\begin{proof}
The proof steps are similar to those of Theorem $\ref{LP_Equivalences-1}%
$.\ \medskip

$(\implies):$ Let $\left(
\begin{array}
[c]{c}%
w^{\ast}\\
x^{\ast}%
\end{array}
\right)  \in Q_{0}$ be an optimal solution for \textit{Problem} $LP_{0}.$ In
other words, let the following be true:%
\begin{equation}
\forall\left(
\begin{array}
[c]{c}%
w^{\dagger}\\
x^{\dagger}%
\end{array}
\right)  \in Q_{0},\text{ }\left(  \mathbf{0}^{t}\cdot w^{\dag}+\widetilde{c}%
^{t}x^{\dag}\right)  =\widetilde{c}^{t}x^{t}\geq\left(  \mathbf{0}^{t}\cdot
w^{\ast}+\widetilde{c}^{t}x^{\ast}\right)  =\widetilde{c}^{t}x^{\ast
}.\label{Thm_Eq2(1)}%
\end{equation}
Let $y^{\ast}\in F_{2}(w^{\ast},x^{\ast}).$ Assume $\left(
\begin{array}
[c]{c}%
w^{\ast}\\
x^{\ast}\\
y^{\ast}%
\end{array}
\right)  $\ is not optimal for $Q_{2}.$ In other words, assume the follwing is
true:
\begin{align}
\exists\left(
\begin{array}
[c]{c}%
w^{\dagger}\\
x^{\dagger}\\
y^{\dagger}%
\end{array}
\right)  \in Q_{2}: &  \left(  \mathbf{0}^{t}\cdot w^{\dag}+\widetilde{c}%
^{t}x^{\dag}+\mathbf{0}^{t}\cdot y^{\dag}\right)  =\widetilde{c}^{t}%
x^{t}\nonumber\\[0.03in]
&  <\left(  \mathbf{0}^{t}\cdot w^{\ast}+\widetilde{c}^{t}x^{\ast}%
+\mathbf{0}^{t}\cdot y^{\ast}\right)  =\widetilde{c}^{t}x^{\ast}%
.\label{Thm_Eq2(2)}%
\end{align}
By Assumption $\ref{Projection_Assumptions}.2$,
\begin{equation}
\left(
\begin{array}
[c]{c}%
w^{\dagger}\\
x^{\dagger}\\
y^{\dagger}%
\end{array}
\right)  \in Q_{2}\implies\left(
\begin{array}
[c]{c}%
w^{\dagger}\\
x^{\dagger}%
\end{array}
\right)  \in Q_{0}.\label{Thm_Eq2(3)}%
\end{equation}
Hence, (\ref{Thm_Eq2(2)}) contradicts (\ref{Thm_Eq2(1)}). Hence, we must have:%
\begin{align}
&  \left(
\begin{array}
[c]{c}%
w^{\ast}\\
x^{\ast}%
\end{array}
\right)  \text{ optimal for \textit{Problem} }LP_{0}\implies
\nonumber\\[0.03in]
&  \left(
\begin{array}
[c]{c}%
w^{\ast}\\
x^{\ast}\\
y^{\ast}%
\end{array}
\right)  \text{ optimal for \textit{Problem} }LP_{2}\text{ for each }y^{\ast
}\in F_{2}(w^{\ast},x^{\ast}).\label{Thm_Eq2(4)}%
\end{align}
\smallskip

$(\impliedby)$: Assume $\left(
\begin{array}
[c]{c}%
w^{\ast}\\
x^{\ast}\\
y^{\ast}%
\end{array}
\right)  \in Q_{2}$ is optimal for \textit{Problem} $LP_{2}$ for each
$y^{\ast}\in F_{2}(w^{\ast},x^{\ast}).$ In other words, assume we have:%
\begin{align}
\forall y^{\ast}\in F_{2}(w^{\ast},x^{\ast}),\text{ }\forall\left(
\begin{array}
[c]{c}%
w^{\dagger}\\
x^{\dagger}\\
y^{\dagger}%
\end{array}
\right)  \in Q_{2},\text{ } &  \left(  \mathbf{0}^{t}\cdot w^{\dag
}+\widetilde{c}^{t}x^{\dag}+\mathbf{0}^{t}\cdot y^{\dag}\right)
=\widetilde{c}^{t}x^{t}\nonumber\\[0.03in]
&  \geq\left(  \mathbf{0}^{t}\cdot w^{\ast}+\widetilde{c}^{t}x^{\ast
}+\mathbf{0}^{t}\cdot y^{\ast}\right)  =\widetilde{c}^{t}x^{\ast
}.\label{Thm_Eq2(6)}%
\end{align}
Now, assume $\left(
\begin{array}
[c]{c}%
w^{\ast}\\
x^{\ast}%
\end{array}
\right)  $ is not optimal for \textit{Problem} $LP_{0}$. Then, we must have:
\begin{equation}
\exists\left(
\begin{array}
[c]{c}%
w^{\dagger}\\
x^{\dagger}%
\end{array}
\right)  \in Q_{0}:\left(  \mathbf{0}^{t}\cdot w^{\dag}+\widetilde{c}%
^{t}x^{\dag}\right)  =\widetilde{c}^{t}x^{t}<\left(  \mathbf{0}^{t}\cdot
w^{\ast}+\widetilde{c}^{t}x^{\ast}\right)  =\widetilde{c}^{t}x^{\ast
}.\label{Thm_Eq2(7)}%
\end{equation}
By Assumption $\ref{Projection_Assumptions}.2$,
\begin{equation}
\left(
\begin{array}
[c]{c}%
w^{\dagger}\\
x^{\dagger}\\
y^{\dagger}%
\end{array}
\right)  \in Q_{2}\implies\left(
\begin{array}
[c]{c}%
w^{\dagger}\\
x^{\dagger}%
\end{array}
\right)  \in Q_{0}.\label{Thm_Eq2(8)}%
\end{equation}
Hence, statement (\ref{Thm_Eq2(7)}) contradicts statement (\ref{Thm_Eq2(6)}).
Hence, we must have:%
\begin{align}
&  \left(
\begin{array}
[c]{c}%
w^{\ast}\\
x^{\ast}\\
y^{\ast}%
\end{array}
\right)  \text{ optimal for \textit{Problem} }LP_{2}\text{ for each }y^{\ast
}\in F_{2}(w^{\ast},x^{\ast})\implies\nonumber\\[0.03in]
&  \left(
\begin{array}
[c]{c}%
w^{\ast}\\
x^{\ast}%
\end{array}
\right)  \text{ optimal for \textit{Problem} }LP_{0}.\label{Thm_Eq2(9)}%
\end{align}
\medskip
\end{proof}

Our central thesis will now be formally stated. \ 

\begin{theorem}
\label{LP_Equivalences-0}Assume that the extreme points of \textit{Polytope}
$Q_{0}$ are in one-to-one correspondence with the extreme points of the LAP
polytope, $\mathcal{A}_{n}.$ Let $\widehat{w}\in Ext(\mathcal{A}_{n})$ be the
extreme point of $\mathcal{A}_{n}$ associated with $\widehat{x}\in
\mathbb{R}^{\kappa}$ which is such that $\left(
\begin{array}
[c]{c}%
\widehat{w}\\
\widehat{x}%
\end{array}
\right)  \in Ext(Q_{0}).$ Let $\tau(\widehat{w})$ be the TSP tour
corresponding to $\widehat{w}.$ Assume $\widetilde{c}$ is such that
$\widetilde{c}^{t}\widehat{x}$ is equal to the cost of the travels involved in
$\tau(\widehat{w}).$ Then, \textit{Problem} $LP_{0}$ correctly solves the TSP
optimization problem, irrespective of $Q_{0}$ projecting to \textquotedblleft%
\textit{the} TSP polytope\textquotedblright\ ($\mathcal{P}_{n}$) or not.
\end{theorem}

\begin{proof}
The (premised) one-to-one correspondence between $Ext(\mathcal{A}_{n})$ and
$Ext(Q_{0})$ and the one-to-one correspondence between $Ext(\mathcal{A}_{n})$
and TSP tours imply (by transitivity):%
\begin{equation}
\text{There exists a one-to-one correspondence between }Ext(Q_{0})\text{ and
TSP tours.} \label{LP_Eq_Thm0(1)}%
\end{equation}
Also, for each $\left(
\begin{array}
[c]{c}%
\widehat{w}\\
\widehat{x}%
\end{array}
\right)  \in Ext(Q_{0})$, $\widehat{w}\in Ext(\mathcal{A}_{n})$, and the
unique TSP tour, $\tau(\widehat{w})$, associated with $\widehat{w}$ must also
be associated with $\left(
\begin{array}
[c]{c}%
\widehat{w}\\
\widehat{x}%
\end{array}
\right)  $. Hence, by premise, $\left(
\begin{array}
[c]{c}%
\mathbf{0}\\
\widetilde{c}%
\end{array}
\right)  ^{t}\cdot\left(
\begin{array}
[c]{c}%
\widehat{w}\\
\widehat{x}%
\end{array}
\right)  =\widetilde{c}^{t}\widehat{x}$ correctly accounts the total cost of
the travels involved in $\tau(\widehat{w})$, for each $\left(
\begin{array}
[c]{c}%
\widehat{w}\\
\widehat{x}%
\end{array}
\right)  \in Ext(Q_{0})$. The theorem follows directly from the combination of
this and Theorem \ref{LP_Equivalences-2}.
\end{proof}

\begin{corollary}
A finding that \textquotedblleft\textit{the} TSP polytope\textquotedblright%
\ does not have a polynomial-sized EF does not imply that there does not exist
a polynomial-sized LP which can solve the TSP optimization problem. \ 
\end{corollary}

\begin{remark}
\label{Exact_EF_of_AP_Rmk}If \textit{Polytope} $Q_{0}$ is integral in addition
to its extreme points being in one-to-one correspondence with the extreme
points of \textit{Polytope} $\mathcal{A}_{n}$, then there exist a polynomially
bounded number of $x$-variables for which several alternate ways for
\textquotedblleft constructing\textquotedblright\ costs $\widetilde{c}$ so
that TSP tour costs are correctly captured. This is shown in the next theorem.
In order not to detract from the focus of this paper, we will leave the
(difficult) issue of fnding extension constraints of $Q_{0}$ (i.e., finding
$\mathbf{A}$\textbf{, }$\mathbf{B}$\textbf{\textit{,}} and $\mathbf{b}$) so
that $Q_{0}$ is an exact EF of $\mathcal{A}_{n}$ as an open topic for research.
\end{remark}

\begin{theorem}
\label{Applied_Costs_Thm}For each triplet of nodes of the TSPAG, $\left(
\lceil i,p\rfloor,\lceil j,r\rfloor,\lceil k,s\rfloor\right)  $, define a
variable $x_{\lceil i,p\rfloor\lceil j,r\rfloor\lceil k,s\rfloor}$ and
$\widetilde{c}\in\mathbb{R}^{m^{6}}$ based on the TSP travel costs, $d$, as
follows:%
\[
\widetilde{c}_{\lceil i,p\rfloor\lceil j,r\rfloor\lceil k,s\rfloor}:=\left\{
\begin{array}
[c]{l}%
d_{0i}+d_{ij}+d_{jk}\text{ \ \ \ if \ (}p=1;\text{ }r=2;\text{ }s=3\text{);}\\
\\
d_{jk}+d_{k0}\text{ \ \ \ if \ (}p=1;\text{ }r=m-1;\text{ }s=m\text{);}\\
\\
d_{jk}\text{ \ \ if \ (}p=1;\text{ }3\leq r\leq m-2;\text{ }s=r+1\text{);}\\
\\
0\text{ \ \ \ Otherwise.}%
\end{array}
\right.
\]

Assume the following are valid for $Q_{0}$: \ 

\begin{description}
\item $(i)$ $x_{\lceil i,p\rfloor\lceil j,r\rfloor\lceil k,s\rfloor}>0\implies
p<r<s$ and $i\neq j\neq k$;

\item $(ii)$ $x_{\lceil i,p\rfloor\lceil j,r\rfloor\lceil k,s\rfloor}\leq
min(w_{\lceil i,p\rfloor},w_{\lceil j,r\rfloor},w_{\lceil k,s\rfloor});$

\item $(iii)$ $w_{\lceil i,p\rfloor}+w_{\lceil j,r\rfloor}+w_{\lceil
k,s\rfloor}-x_{\lceil i,p\rfloor\lceil j,r\rfloor\lceil k,s\rfloor}\leq2.$
\end{description}

Then, for each $\left(
\begin{array}
[c]{c}%
\widehat{w}\\
\widehat{x}%
\end{array}
\right)  \in Ext(Q_{0}),$%
\[
\mathcal{V}\left(  w,x\right)  :=\left(
\begin{array}
[c]{c}%
\mathbf{0}\\
\widetilde{c}%
\end{array}
\right)  ^{t}\cdot\left(
\begin{array}
[c]{c}%
\widehat{w}\\
\widehat{x}%
\end{array}
\right)  =\sum\limits_{(\lceil i,p\rfloor,\lceil j,r\rfloor,\lceil
k,s\rfloor)\in N^{3}}\widetilde{c}_{\lceil i,p\rfloor\lceil j,r\rfloor\lceil
k,s\rfloor}\widehat{x}_{\lceil i,p\rfloor\lceil j,r\rfloor\lceil k,s\rfloor}%
\]
correctly accounts the total cost of the travels involved in the TSP tour
associated with $\widehat{w}$ and $\left(
\begin{array}
[c]{c}%
\widehat{w}\\
\widehat{x}%
\end{array}
\right)  .$
\end{theorem}

\begin{proof}
Let $\left(
\begin{array}
[c]{c}%
\widehat{w}\\
\widehat{x}%
\end{array}
\right)  \in Ext(Q_{0}).$ Then, by premise, $\widehat{w}\in Ext(\mathcal{A}%
_{n}).$ Hence, letting the positive entries of $\widehat{w}$ be $\widehat{w}%
_{u_{1},1},\widehat{w}_{u_{2},2},\ldots,\widehat{w}_{u_{m},m},$ we must have:
\begin{align}
&  \forall(p,q)\in S^{2}:p\neq q,\text{ }u_{p}\neq u_{q};\text{ and}%
\nonumber\\
&  \forall\lceil v,q\rfloor\in N,\text{ }\widehat{w}_{vq}=\left\{
\begin{array}
[c]{l}%
1\text{ \ \ if }v=u_{q}\\
\\
0\text{ \ \ otherwise.}%
\end{array}
\right.  . \label{A_Costs_Thm(1)}%
\end{align}
Hence, the (unique) TSP tour corresponding to $\widehat{w}$ can be stated as
$\tau$($\widehat{w}):=$ $0\rightarrow u_{1}^{k}\rightarrow\ldots\rightarrow
u_{m}^{k}\rightarrow0$ (as discussed earlier in section
\ref{LAP_Representation_Subsection}). The total cost of the travels involved
in $\tau(\widehat{w})$ is:
\begin{equation}
TCost=d_{0,u_{1}}+d_{u_{m},0}+{\displaystyle\sum\limits_{q=1}^{m-1}}%
d_{u_{q},u_{q+1}}. \label{ObjCostsProof(b)}%
\end{equation}

To account the total cost involved in $\widehat{x},$ observe that statement
($\ref{A_Costs_Thm(1)}$) and conditions $(i)-(iii)$ imply:%
\begin{align}
&  \forall(\lceil v_{p},p\rfloor,\lceil v_{r},r\rfloor,\lceil v_{s}%
,s\rfloor)\in N^{3},\nonumber\\
&  \widehat{x}_{\lceil v_{p},p\rfloor\lceil v_{r},r\rfloor\lceil
v_{s},s\rfloor}=\left\{
\begin{array}
[c]{l}%
1\text{ \ \ if }v_{p}=u_{p},\text{ }v_{r}=u_{r},\text{ }v_{s}=u_{s},\text{
}p<r<s;\\
\\
0\text{ \ \ otherwise.}%
\end{array}
\right.  \label{A_Costs_Thm(2)}%
\end{align}

Hence, focusing on the components of $\widehat{x}$ and using
(\ref{A_Costs_Thm(2)}), the costs incurred by $\widehat{x}$ are as follows:
\bigskip%

\begin{tabular}
[c]{l}%
\ \ \ \ \ \ \
\end{tabular}%
\begin{tabular}
[c]{|p{2.25in}|p{2.25in}|}\hline
\multicolumn{1}{|c|}{} & \multicolumn{1}{|c|}{}\\
\multicolumn{1}{|c|}{Component, $\widehat{x}_{\lceil u_{p},p\rfloor\lceil
u_{r},r\rfloor\lceil u_{s},s\rfloor}$} & \multicolumn{1}{|c|}{Cost,
$\widetilde{c}_{\lceil u_{p},p\rfloor\lceil u_{r},r\rfloor\lceil
u_{s},s\rfloor}$}\\
\multicolumn{1}{|c|}{} & \multicolumn{1}{|c|}{}\\\hline
\multicolumn{1}{|p{2.25in}|}{$p=1;$ $r=2;$ $s=3$} & $d_{0,u_{1}}%
+d_{u_{1},u_{2}^{k}}+d_{u_{2},u_{3}}$\\\hline
\multicolumn{1}{|p{2.25in}|}{$p=1;$ $r=m-1;$ $s=m$} & $d_{u_{m-1},i_{m}%
}+d_{u_{m},0}$\\\hline
\multicolumn{1}{|p{2.25in}|}{$p=1;$ $r=3;$ $s=4$} & $d_{u_{3},u_{4}}$\\\hline
\multicolumn{1}{|p{2.25in}|}{$\vdots$} & $\vdots$\\\hline
\multicolumn{1}{|p{2.25in}|}{$p=1;$ $r=m-2;$ $s=m-1$} & $d_{u_{m-2},u_{m-1}}%
$\\\hline
Total cost attached to $\left(
\begin{array}
[c]{c}%
\widehat{w}\\
\widehat{x}%
\end{array}
\right)  =$ & $d_{0,u_{1}}+d_{u_{m},0}+{\displaystyle\sum\limits_{q=1}^{m-1}%
}d_{u_{q},u_{q+1}}$\\\hline
\end{tabular}
\medskip\medskip

Comparing the results above, we observe that the total cost associated with
$\left(
\begin{array}
[c]{c}%
\widehat{w}\\
\widehat{x}%
\end{array}
\right)  $ is indeed equal to that of the TSP tour represented by
$\widehat{w},$ $\tau(\widehat{w})$. \ \ \medskip
\end{proof}

Hence,\ in particular, if $Q_{0}$ can be described by a polynomial-sized set
of constraints in the space of the $x$-variables specified in Theorem
\ref{Applied_Costs_Thm} in such a way that it is integral and has extreme
points which are in one-to-one correspondence with those of the LAP polytope
$\mathcal{A}_{n}$, then the TSP optimization problem can be solved as a
polynomial sized LP without recourse to \textquotedblleft\textit{the} TSP
polytope\textquotedblright\ $\mathcal{P}_{n},$ and thereby side-stepping the
existing \textit{barriers}.\medskip

\section{Conclusions\label{Conclusions_Section}}

We have provided conditions under which the TSP optimization problem may be
solvable without recourse to \textquotedblleft\textit{the} TSP
polytope.\textquotedblright\ The possibility of these conditions is not ruled
out by any existing result, and hence subject to this,\ we have shown that a
finding that \textquotedblleft\textit{the} TSP polytope\textquotedblright%
\ does not have a polynomial-sized \textit{exact} extended formulation (EF)
does not imply that it is \textquotedblleft impossible\textquotedblright\ for
a linear program (polynomial-sized or not) to solve the TSP\ optimization
problem. Our demonstrations are based on a time-dependent/Assignment Problem
representation of TSP tours. We have shown that provided an EF
(polynomial-sized or not) of the Linear Assignment Problem (LAP) polytope is
\textit{exact} and that TSP\ tour costs can be accurately accounted through
linear functions of its descriptive variables, that EF would solve the TSP
optimization problem, irrespective of whether a further EF of it in the space
of the city-to-city (\textit{travel leg}) variables projects to
\textquotedblleft\textit{the} TSP polytope\textquotedblright\ or not. We have
also proposed variables which are auxiliary to the natural descriptive
variables of the LAP polytope and in terms of which TSP tour costs can be
correctly captured using linear functions. Also, because the combinatorial
configurations of many of the other NP-Complete problems can be modeled as a
LAP with \textquotedblleft side-constraints,\textquotedblright\ our
developments are directly applicable to many of the other NP-Complete problems.

\begin{center}
{\huge Appendix: Tables of notations\bigskip}

\bigskip\bigskip%

\begin{tabular}
[c]{ll}\hline
\multicolumn{1}{|l}{\textbf{Notation}} & \multicolumn{1}{|l|}{\textbf{Meaning}%
}\\\hline
\multicolumn{1}{|l}{$n$} & \multicolumn{1}{|l|}{Number of TSP cities.}\\\hline
\multicolumn{1}{|l}{$m:=n-1$} & \multicolumn{1}{|l|}{Numbers of
\textit{levels} and \textit{stages} respectively, of the TSPAG}\\\hline
\multicolumn{1}{|l}{$C:=\left\{  0,\ldots,m\right\}  $} &
\multicolumn{1}{|l|}{Set of cities}\\\hline
\multicolumn{1}{|l}{$\forall(i,j)\in C^{2},$ $d_{ij}$\ } &
\multicolumn{1}{|l|}{Cost of traveling from city $i$ to city $j$}\\\hline
\multicolumn{1}{|l}{$\ell:=n\times m$} & \multicolumn{1}{|l|}{Number of arcs
in the \textit{TSP graph}}\\\hline
\multicolumn{1}{|l}{$\mathcal{P}_{n}$} &
\multicolumn{1}{|l|}{\textquotedblleft\textit{the} TSP
polytope\textquotedblright\ defined over $n $ cities}\\\hline
\multicolumn{1}{|l}{$L:=\{1,\ldots,m\}$} & \multicolumn{1}{|l|}{Index set for
the \textit{levels} of the TSPAG}\\\hline
\multicolumn{1}{|l}{$S:=\{1,\ldots,m\}$} & \multicolumn{1}{|l|}{Index set for
the \textit{stages} of the TSPAG}\\\hline
\multicolumn{1}{|l}{$N:=\left\{  (l,s)\in(L,S)\right\}  $} &
\multicolumn{1}{|l|}{Set of nodes of the TSPAG}\\\hline
\multicolumn{1}{|l}{$\mathcal{A}_{n}$} & \multicolumn{1}{|l|}{The LAP polytope
defined over the TSPAG associated with a $n$-city TSP}\\\hline
\multicolumn{1}{|l}{$(A)^{t}$} & \multicolumn{1}{|l|}{Transpose of matrix $A$%
}\\\hline
\multicolumn{1}{|l}{$Ext(B)$\ } & \multicolumn{1}{|l|}{Set of extreme points
of polyhedron $B$}\\\hline
\multicolumn{1}{|l}{$\pi_{x}(B)$} & \multicolumn{1}{|l|}{Projection of
polyhedron $B$ on the space of vector $x$}\\\hline
\multicolumn{1}{|l}{$\mathbb{R}$} & \multicolumn{1}{|l|}{Set of real
numbers}\\\hline
\multicolumn{1}{|l}{$\mathbb{R}_{\mathbb{\geq}0}$} & \multicolumn{1}{|l|}{Set
of nonnegative real numbers}\\\hline
\multicolumn{1}{|l}{$\mathbb{N}$\ } & \multicolumn{1}{|l|}{Set of natural
numbers}\\\hline
& \\
\multicolumn{2}{c}{\textbf{Table A.1: List of general notations}}\\
&
\end{tabular}

\bigskip

\bigskip%

\begin{tabular}
[c]{ll}\hline
\multicolumn{1}{|l}{\textbf{Notation}} & \multicolumn{1}{|l|}{\textbf{Meaning}%
}\\\hline
\multicolumn{1}{|c}{$\forall w\in\mathcal{A}_{n},$ $F_{0}(w):=\left\{
x\in\mathbb{R}^{k}:\left(
\begin{array}
[c]{c}%
w\\
x
\end{array}
\right)  \in Q_{0}\right\}  $} & \multicolumn{1}{|c|}{Set of $x$-vectors such
that $\left(
\begin{array}
[c]{c}%
w\\
x
\end{array}
\right)  $ is feasible for $Q_{0}$}\\\hline
\multicolumn{1}{|c}{$\forall w\in\mathcal{A}_{n},$ $F_{1}(w):=\left\{
y\in\mathbb{R}^{\ell}:\left(
\begin{array}
[c]{c}%
w\\
y
\end{array}
\right)  \in Q_{1}\right\}  $} & \multicolumn{1}{|c|}{Set of $y$-vectors such
that $\left(
\begin{array}
[c]{c}%
w\\
y
\end{array}
\right)  $ is feasible for $Q_{1}$}\\\hline
\multicolumn{1}{|l|}{$\forall w\in\mathcal{A}_{n},$ $\forall x\in
\mathbb{R}^{k}:\left(
\begin{array}
[c]{c}%
w\\
y
\end{array}
\right)  \in Q_{0},$} & \multicolumn{1}{c|}{}\\
\multicolumn{1}{|l|}{$F_{2}(w,x):=\left\{  y\in\mathbb{R}^{\ell}:\left(
\begin{array}
[c]{c}%
w\\
x\\
y
\end{array}
\right)  \in Q_{2}\right\}  $} & \multicolumn{1}{l|}{Set of $y$-vectors such
that $\left(
\begin{array}
[c]{c}%
w\\
x\\
y
\end{array}
\right)  $ is feasible for $Q_{2}$}\\\hline
& \\
\multicolumn{2}{c}{\textbf{Table A.2: Feasible sets definitions}}\\
& \\
&
\end{tabular}

\end{center}

\end{document}